\documentclass[%
 reprint,
 superscriptaddress,
 amsmath,amssymb,
 aps,
 prb,
 floatfix,
]{revtex4-2}

\usepackage{graphicx}
\usepackage{dcolumn}
\usepackage{bm}

\usepackage{fontawesome5}
\usepackage{xcolor}
\usepackage{comment}
\usepackage{booktabs}
\usepackage{threeparttable}
\usepackage{amsmath}

\DeclareMathOperator*{\argmin}{arg\!\min}

\makeatletter
\newcommand{\github}[1]{%
   \href{#1}{\faGithubSquare}%
}

\definecolor{revision}{RGB}{0,0,0}

\begin{document}

\title{Elucidating proximity magnetism through polarized neutron reflectometry and machine learning}

\author{Nina Andrejevic}
\thanks{These authors contributed equally.}
\affiliation{Quantum Measurement Group, Massachusetts Institute of Technology, Cambridge, MA 02139, USA}
\affiliation{Department of Materials Science and Engineering, Massachusetts Institute of Technology, Cambridge, MA 02139, USA}
\author{\!\!\!$^{,\ \! \dagger}$ Zhantao Chen}
\thanks{These authors contributed equally.}
\affiliation{Quantum Measurement Group, Massachusetts Institute of Technology, Cambridge, MA 02139, USA}
\affiliation{Department of Mechanical Engineering, Massachusetts Institute of Technology, Cambridge, MA 02139, USA}
\author{Thanh Nguyen}
\affiliation{Quantum Measurement Group, Massachusetts Institute of Technology, Cambridge, MA 02139, USA}
\affiliation{Department of Nuclear Science and Engineering, Massachusetts Institute of Technology, Cambridge, MA 02139, USA}
\author{Leon Fan}
\affiliation{Department of Electrical Engineering and Computer Science, Massachusetts Institute of Technology, Cambridge, MA 02139, USA}
\author{Henry Heiberger}
\affiliation{Department of Electrical Engineering and Computer Science, Massachusetts Institute of Technology, Cambridge, MA 02139, USA}
\author{Ling-Jie Zhou}
\affiliation{Department of Physics, The Pennsylvania State University, University Park, PA 16802, USA}
\author{Yi-Fan Zhao}
\affiliation{Department of Physics, The Pennsylvania State University, University Park, PA 16802, USA}
\author{Cui-Zu Chang}
\affiliation{Department of Physics, The Pennsylvania State University, University Park, PA 16802, USA}
\author{Alexander Grutter}
\affiliation{National Institute of Standards and Technology, Center for Neutron Research, Gaithersburg, MD 20899, USA}
\author{Mingda Li}
\thanks{Corresponding author.\vspace{0.05cm} \\\href{mailto:nina12@mit.edu}{nina12@mit.edu}
\\\href{mailto:mingda@mit.edu}{mingda@mit.edu}}
\affiliation{Quantum Measurement Group, Massachusetts Institute of Technology, Cambridge, MA 02139, USA}
\affiliation{Department of Nuclear Science and Engineering, Massachusetts Institute of Technology, Cambridge, MA 02139, USA}

\date{\today}

\begin{abstract}
\noindent \textbf{Abstract} \\
Polarized neutron reflectometry is a powerful technique to interrogate the structures of multilayered magnetic materials with depth sensitivity and nanometer resolution. However, reflectometry profiles often inhabit a complicated objective function landscape using traditional fitting methods, posing a significant challenge to parameter retrieval. In this work, we develop a data-driven framework to recover the sample parameters from polarized neutron reflectometry data with minimal user intervention. We train a variational autoencoder to map reflectometry profiles with moderate experimental noise to an interpretable, low-dimensional space from which sample parameters can be extracted with high resolution. We apply our method to recover the scattering length density profiles of the topological insulator-ferromagnetic insulator heterostructure Bi$_2$Se$_3$/EuS exhibiting proximity magnetism, in good agreement with the results of conventional fitting. We further analyze a more challenging reflectometry profile of the topological insulator-antiferromagnet heterostructure (Bi,Sb)$_2$Te$_3$/Cr$_2$O$_3$ and identify possible interfacial proximity magnetism in this material. We anticipate the framework developed here can be applied to resolve hidden interfacial phenomena in a broad range of layered systems.
\github{https://github.com/ninarina12/ML_PNR}
\\


\end{abstract}

\maketitle

Neutron reflectometry facilitates structural characterization of multilayered materials by probing their nuclear and magnetic depth profiles at device-relevant spatial scales, enabling the study of hidden interfaces in a broad range of nanostructured and thin film systems \cite{fitzsimmons2004neutron,lauter2007neutron,fitzsimmons2007pinned,bennett2016giant,gilbert2016structural,theis2020self,need2020magnetic,keunecke2020high,liu2021ferroelectric,bhatnagar2021differentiation,wang2021optically}. Leveraging the interaction of spin-polarized neutrons with magnetic moments, polarized neutron reflectometry (PNR) is particularly well-suited to detecting magnetic interfacial phenomena \cite{ankner1999polarized,majkrzak2006polarized,toperverg2015polarized,nichols2016emerging,zhan2019probing} such as the magnetic proximity effect. Proximity coupling to a magnetic material induces magnetic order near the interface of an otherwise non-magnetic system, making it a promising pathway for magnetizing topological insulators (TIs) without introducing magnetic dopants \cite{bhattacharyya2021recent,vobornik2011magnetic,eremeev2013magnetic,lang2014proximity,lee2014magnetic,li2015magnetic,liu2015enhancing,katmis2016high,he2017tailoring,che2018proximity,koren2018magnetic,he2018exchange,he2018topological,hou2019magnetizing,akiyama2019direct,watanabe2019quantum,pan2020observation,li2015proximity}. This opens up the possibility of realizing emergent phenomena such as the quantum anomalous Hall effect \cite{tokura2019magnetic,yu2010quantized,kou2014scale,kou2015magnetic,mogi2019large} or axion insulator state \cite{mogi2017magnetic,mogi2017tailoring,xiao2018realization} at room temperature and advancing TI-based device applications. More recently, the realization of proximity magnetism in van der Waals heterostructures \cite{liang2017magnetic,karpiak2019magnetic,tong2019magnetic,behera2019proximity,huang2020emergent,zhao2020magnetic,tang2020magnetic,zhong2020layer,dayen2020two,zhang2020controllable,zhang2020proximity,bora2021magnetic} presents new opportunities to engineer atomically thin devices with novel functionalities, and at the same time highlights an increasing need for precise characterization of interfacial effects at subnanometer length scales. Thus, accurate quantitative analysis of magnetic structural information obtained by PNR is critical to resolving important interfacial effects within a broad range of materials systems.

However, subtle interfacial phenomena such as proximity magnetism can be difficult to single out from bulk contributions to the PNR signature. The composition and magnetic profiles of a sample are typically expressed in terms of the nuclear and magnetic scattering length densities (SLD), which can be recovered from reflectometry measurements by fitting the data to a candidate model. This traditionally involves building a theoretical model of the experimental system in terms of structural parameters -- density, thickness, interface roughness, and magnetization -- of the constituent layers, and simulating the associated reflectometry profile using the methods of Parratt recursion \cite{parratt1954surface} or the Abeles matrix formalism \cite{abeles1948propagation}. However, due to information loss about the phase of the reflected neutrons, different SLD profiles can generate highly similar reflectivities, leading to a complicated cost landscape between the theoretical and experimental profiles with potentially many local minima. Thus, parameter refinement often demands expert insight to identify a suitable starting point and adequately constrain the parameter space of the model. Methods to resolve the phase ambiguity through carefully designed experiments \cite{sivia1991novel,de1995retrieval,pleshanov1999polarized,o2002pinpointing,durant2021optimising} have also been proposed. More generally, additional insights from X-ray diffraction, transmission electron microscopy, and/or bulk magnetometry are required, as well as selection of appropriate models \cite{sivia1998bayesian,mccluskey2020general} and optimization methods, which can play a critical role in steering the refinement process. For example, many existing neutron and X-ray reflectivity refinement programs (e.g. GenX \cite{bjorck2007genx}, Refl1D \cite{maranville2017interactive}, StochFit \cite{danauskas2008stochastic}), implement stochastic optimization methods such as differential evolution, simulated annealing, or stochastic tunneling, to better manage multiple local minima. More recently, machine learning-guided fitting approaches have been introduced to both improve and automate parameter retrieval from neutron reflectometry data with promising results \cite{greco2019fast,mironov2021towards,loaiza2021towards,doucet2021machine,aoki2021deep}. However, elucidating subtle interfacial phenomena such as magnetic proximity from PNR signatures remains a significant challenge.

In this work, we develop an alternate, data-driven framework to retrieve the sample parameters of candidate proximity-coupled systems from their PNR profiles with minimal user intervention. Using a variational autoencoder (VAE), we map reflectometry profiles simulated from a broad range of candidate physical parameters to a low-dimensional latent space from which the true sample parameters can be readily obtained. The decoded profiles directly inform the suitability of the parameter search space through the reconstruction quality and are robust to moderate perturbation of the input reflectivities emulating experimental noise. Importantly, we find that the latent mapping naturally bypasses the issue of multiple local minima and is both well-organized and visually interpretable in terms of the physical parameters. Thus, the latent representation can further be used to automatically refine the parameter search space for poorly reconstructed profiles. We evaluate our model on its ability to recover the sample parameters of a well-studied TI-ferromagnetic (FM) insulator heterostructure, Bi$_2$Se$_3$/EuS, exhibiting proximity magnetism. Our model predictions are found to be consistent with the results of traditional fitting methods and at the same time require no expert insight for parameter initialization or refinement. We conclude by applying our model to analyze a more challenging PNR profile of the TI-antiferromagnet (AFM) heterostructure, (Bi,Sb)$_2$Te$_3$/Cr$_2$O$_3$, which points to possible proximity magnetism at the resolution limit.

{\color{revision}\section*{Background}}

\begin{figure*}[t]
    \centering
    \includegraphics[width=0.9\linewidth]{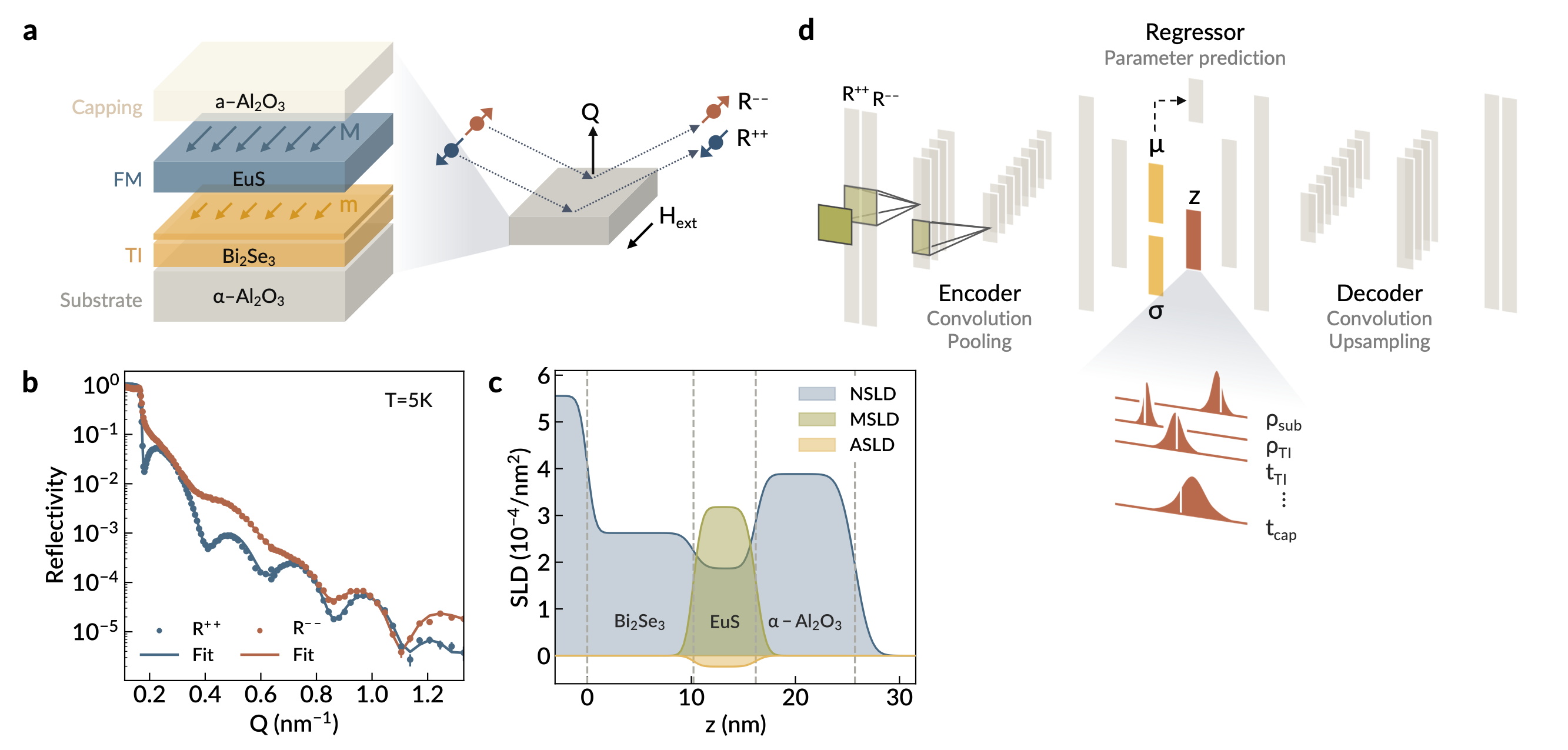}
    \caption{\textbf{VAE with regression for PNR data analysis.} \textbf{a.} Schematic illustration of the proximity-coupled Bi$_2$Se$_3$/EuS system. A depiction of the spin-polarized neutron reflectometry experiment under an externally applied magnetic field is shown at right. \textbf{b.} Reflectometry profile of the heterostructure in \textbf{a} measured at T = 5 K. Solid lines correspond to {\color{revision}a representative} fit obtained using the GenX paramter refinement program. Error-bars represent ±1 standard deviation. \textbf{c.} Nuclear, magnetic, and absorption SLD profiles corresponding to the fit in \textbf{b}. \textbf{d.} Schematic illustration of the VAE architecture used for PNR parameter retrieval.}
    \label{fig:Fig1}
\end{figure*}

{\color{revision}Polarized neutron reflectometry measures the spin-dependent specular reflection of an incident neutron beam from the surface of a magnetic thin film. The reflectometry profile, $R(Q)$, is a function of the wave vector transfer $Q = 4 \pi \sin{\theta}/\lambda$, where $\theta$ is the angle of reflection and $\lambda$ is the wavelength of the neutron. In the first Born approximation, the reflectivities of the neutron spin non-flip channels, $R^{++}(Q)$ and $R^{--}(Q)$, are related to the nuclear and magnetic SLDs according to \cite{majkrzak2006polarized}},
\begin{equation*}
    R^{\pm\pm}(Q) = \frac{16 \pi^2}{Q^2}\left| \int_{-\infty}^{\infty}\left(\rho_N(z) \pm \rho_M(z)\cos{\phi}\right)\mathrm{e}^{-iQz}dz \right|^2,
\end{equation*}
{\color{revision}where $\rho_N$ and $\rho_M$ are, respectively, the nuclear and magnetic scattering length densities; $\phi$ is the angle between the magnetization and the neutron polarization; and the superscript $+$ and $-$ denote the neutron spin up and down states, respectively. The coordinate $z$ measures the depth perpendicular to the sample surface.} To develop the VAE-based approach to recover the SLD profiles of candidate proximity-coupled systems, we consider the specific thin film system consisting of a Bi$_2$Se$_3$/EuS heterostructure atop a sapphire ($\alpha-$Al$_2$O$_3$) substrate with an amorphous a-Al$_2$O$_3$ capping layer, as shown in Fig. \ref{fig:Fig1}a. Proximity-induced magnetism has been reported at the interface between the TI Bi$_2$Se$_3$ and EuS, a ferromagnet with a Curie temperature of approximately 16.6 K \cite{katmis2016high,lee2016direct}. The reflectometry profile shown in Fig. \ref{fig:Fig1}b consists of two curves, $R^{++}$ and $R^{--}$, corresponding to the two neutron spin non-flip channels aligned parallel and antiparallel, respectively, to an in-plane external magnetic field $H_\text{ext}$ = 1 T. {\color{revision}Note that $R^{++}$ and $R^{--}$ have been normalized to a maximum value of 1.} In a typical parameter refinement program, the theoretical model is fit simultaneously to both spin channels to obtain the SLD profile of the sample (Fig. \ref{fig:Fig1}c). However, due to the phase ambiguity and large number of fitting parameters, different SLD profiles can produce excellent fits to the measured data. For instance, the fit obtained in Fig. \ref{fig:Fig1}b corresponds to the SLD profile shown in Fig. \ref{fig:Fig1}c, {\color{revision}which proposes a high interface roughness between the Bi$_2$Se$_3$ and EuS films but no discernible proximity effect. Fig. \ref{fig:FigS1} shows three additional fits to the data obtained using the GenX parameter refinement program with different initial populations, which produce mixed results for the relevant parameters.} The objective of the data-driven approach is to retrieve the optimal physical parameters of a target sample from its PNR profile under moderate experimental noise, and to compute a reliable SLD profile from learned sample parameters with minimal influence from common issues in iterative optimization algorithms, such as sensitivity to parameter initialization and stagnation. We find this approach can further inform the suitability of the entire parameter search space, not just the predicted parameters. \\

{\color{revision}\section*{Framework}}
Our approach is based on the VAE \cite{kingma2013auto}, {\color{revision}an unsupervised deep generative model which is trained to recover an input from a low-dimensional encoding by minimizing the associated reconstruction error. The VAE comprises an encoder and decoder network, as shown in Fig. \ref{fig:Fig1}d. The encoder network outputs parameters to a probability density, $q_\theta (z\!\mid\!x)$, parameterized by a set of trainable weights $\theta$, from which the latent features $z$ are sampled. The decoder network outputs the parameters to the probability distribution of the data, $p_\phi (x\!\mid\!z)$, parameterized by a set of weights $\phi$, using the sampled latent features $z$. In contrast to a simple autoencoder, the VAE assumes that the encoded -- or latent -- vector elements are drawn from a prior distribution $p(\textbf{z})$}, which is enforced by an additional regularization term in the loss function,
\begin{equation*}
    \mathcal{L} = -\mathbb{E}_{\textbf{z} \sim \textbf{q}_\theta(\mathbf{z} \mid \mathbf{x})}[\log p_\phi(\mathbf{x}\!\mid\!\mathbf{z})] + \beta D_{KL}(q_\theta(\mathbf{z}\mid\!\mathbf{x})\!\mid\mid\!p(\textbf{z})),
\end{equation*}
where $D_{KL}$ denotes the Kullback-Leibler (KL) divergence, computed between the returned distribution $q_\theta(\textbf{z}|\textbf{x})$ of the latent vector $\textbf{z}$ and the prior distribution $p(\textbf{z})$, and $\beta$ is a hyperparameter regulating the degree of entanglement between the learned latent channels \cite{higgins2016beta}. {\color{revision}The prior distributions are typically modeled as independent unit Gaussians, \textit{i.e.}, $p(z_i) \sim \mathcal{N}(0,1)$, and the approximate posterior $q_\theta(\textbf{z}\mid\!\textbf{x})$ as a Gaussian with mean and variance estimated by the encoder. Additional details of our specific implementation are provided in the Supplementary Information.} By encoding the input as a distribution rather than as a single point, the VAE compels the latent space to be smooth and continuous, with nearby points corresponding to similar reconstructions of the input. Thus, in the context of PNR, the VAE can be considered as a way to map PNR profiles into a well-organized and informative low-dimensional space, as they naturally evolve as a function of a few well-defined structural parameters (Fig. \ref{fig:FigS3}). {\color{revision}The potential advantages of a VAE-based approach to parameter retrieval are further exemplified through a toy example described in the Supplementary Information.} \\

{\color{revision} \subsection*{Network architecture}}
Like {\color{revision}the conventional} fitting programs, the VAE treats the $R^{++}$ and $R^{--}$ channels jointly using a convolutional neural network (CNN) encoder with a {\color{revision} combination of one- and two- dimensional kernels} (Fig. \ref{fig:Fig1}d). The convolutional and pooling layers are followed by a set of fully-connected layers operating on the flattened CNN output, returning the predicted means $\boldsymbol \mu$ and standard deviations $\boldsymbol \sigma$ of the normal distributions from which the latent vector $\textbf{z}$ is sampled. When the latent representation is conditioned on the sample parameters, we can interpret each latent channel as effectively returning a distribution over one parameter's value, illustrated schematically in Fig. \ref{fig:Fig1}d. The mean values of the latent distributions are fed to a simple regressor consisting of a single hidden and activation layer that predicts the physical parameter values. A ReLU activation is used to restrict the predicted values to be non-negative in accordance with the physical parameters. At the same time, the sampled vector $\textbf{z}$ is passed to the decoder, which returns the reconstructed profiles $R^{'++}$ and $R^{'--}$. All three networks -- encoder, decoder, and regressor -- are trained end-to-end by minimizing the total loss function,
\begin{eqnarray*}\label{eqn:loss_full}
    \mathcal{L} = &-&\mathbb{E}_{\textbf{z} \sim \textbf{q}_\theta(\mathbf{z} \mid \mathbf{x})}[\log p_\phi(\mathbf{x}\!\mid\!\mathbf{z})] \\ &+& \beta D_{KL}(q_\theta(\mathbf{z}\mid\!\mathbf{x})\!\mid\mid\!p(\textbf{z})) + \lambda \|\textbf{v} - \textbf{v}'\|^2,
\end{eqnarray*}
where $\textbf{v}$ and $\textbf{v}'$ denote the true and predicted parameter values, respectively, and $\lambda$ is a hyperparameter weighting the contribution of parameter regression to the total loss. {\color{revision}Details regarding the implementation of the loss function and the selected hyperparameters are provided in the Supplementary Information.}\\

\begin{figure*}[t]
    \centering
    \includegraphics[width=0.9\linewidth]{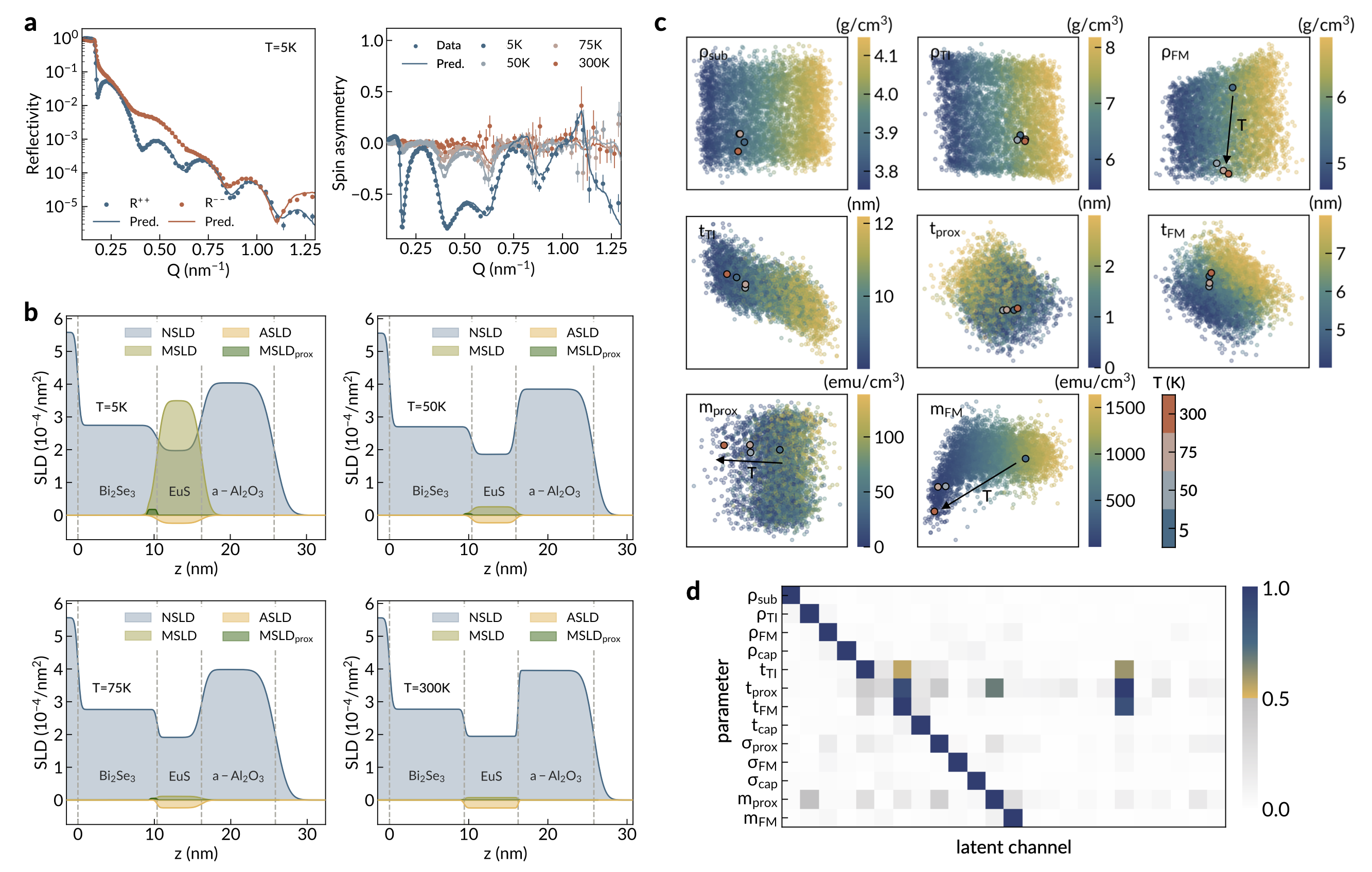}
    \caption{\textbf{VAE performance on experimental data.} \textbf{a.} At left, the decoded PNR profile of the Bi$_2$Se$_3$/EuS heterostructure measured at 5 K. For each channel, points represent experimental data and solid lines are the corresponding reconstruction. At right, the spin asymmetry ($R^{++} -R^{--})/(R^{++} +R^{--}$) calculated from the data (points) and decoded profiles (solid lines) of four PNR measurements taken at different temperatures. Error-bars represent ±1 standard deviation. \textbf{b.} Nuclear (NSLD), magnetic (MSLD), and absorption (ASLD) scattering length density profiles obtained from the regressor predictions for the measurements at the four temperatures in \textbf{a}. {\color{revision}The MSLD contribution from the proximity layer is shown in dark green.} \textbf{c.} Projections of the latent encoding of the test dataset along the latent dimensions with the largest gradients for different sample parameters (density $\rho$, thickness $t$, magnetization $m$), colored by their true values. Outlined points show the latent encoding of the four experimental measurements from \textbf{a}. Specifically, the horizontal and vertical axes correspond to the latent dimensions with the largest and second largest gradient of the target parameter, respectively. \textbf{d.} Parameter entanglement inferred from gradients along each latent channel. Heatmap indicates the relative magnitudes of the gradients of a given parameter with respect to each latent channel. Namely, each cell is colored by the normalized value of $\partial v_{i}/\partial z_{j}$ for the $i$-th parameter and $j$-th latent channel. Normalization maps gradients in each row to lie between 0 and 1.}
    \label{fig:Fig2}
\end{figure*}

{\color{revision} \subsection*{Data preparation}}
To generate the training and development datasets for the neural network model, we used the GenX neutron reflectivity modeling code \cite{bjorck2007genx} to simulate the PNR profiles of {\color{revision}$200,000$} candidate systems of the Bi$_2$Se$_3$/EuS heterostructure. For each example, the constituent layers are parameterized by their density, thickness, roughness, and magnetization, which are sampled uniformly at random over a range of experimentally feasible values (Fig. \ref{fig:FigS4}). Importantly, these parameter ranges can be quite broad around the set of nominal parameter values, and can differ in size for different quantities depending on their level of uncertainty. For example, the parameter ranges for the amorphous capping layer are intentionally broader compared to the TI and FM layer thicknesses that are carefully controlled during growth. Density and magnetization are {\color{revision}expressed} in terms of formula units, which are compatible with the GenX simulation software; however, the final results are converted to conventional units before plotting. The proximity effect is modeled as a thin interfacial layer between the Bi$_2$Se$_3$ and EuS films with a sampled thickness, roughness, and magnetization, and sharing the density value of the neighboring TI film. The proximity layer magnetization is constrained not to exceed the sampled value of the EuS magnetization for any given example. {\color{revision}Additionally, the minimum possible thickness of the proximity layer is set to 2 \AA\, representing a target spatial resolution threshold. Note that the neutron wavelengths for the PNR measurements conducted in this work are on the order of 5 \AA. Examples for which the sampled proximity layer thickness falls below the threshold are simulated without an interfacial layer and are designated as non-proximity-coupled.} The PNR profiles are simulated over the experimentally-accessible $Q$-range from 0.1 to 1.3 nm$^{-1}$, and the intensities normalized to a maximum value of 1. To simulate experimental noise, the generated PNR profiles are randomly perturbed at each $Q$ point by sampling a Gaussian distribution with standard deviation estimated using the errorbars of the corresponding experimental reflectometry profile. {\color{revision}Specific details regarding noise estimation, including selection of the standard deviation and dependence of the results on different noise levels, are provided in the Supplementary Information.} Additionally, the instrument resolution and background are sampled uniformly at random between 0.001 and 0.01 nm$^{-1}$, and 10$^{-8}$ and 10$^{-4}$ on a logarithmic scale, respectively. Since the reflectivity spans nearly eight decades, the base-10 logarithm of the profiles is used as input (output) of the encoder (decoder) to more equitably treat the intensity values. To similarly place the sample parameters on equal footing, the output of the regressor is taken to be the standardized values of the physical parameters. {\color{revision}In particular, the regressor is trained to predict the density, thickness, and roughness of each thin film layer, as well as the magnetization of the ferromagnetic and proximity layers. The substrate density is also predicted by the regressor, but substrate thickness and roughness are excluded from fitting as the substrate is considered macroscopically thick with a relatively uniform surface roughness (approximately 3 \AA\ for the sapphire substrate used in this system). While the instrument resolution and background are not predicted explicitly by the regressor, variations in the data as a function of these parameters can still be captured by the unregularized latent dimensions.} They can thus be regarded as underlying degrees of freedom which may be more complex functions of the latent space. The freedom to choose the number of output quantities, even as the training data reflect variations in the full set of sample and instrument parameters, is one advantage of the machine learning-based approach: It allows one to output only the most relevant quantities, reducing the needed training data volume and neural network size. Additionally, machine learning makes it possible to seek hidden relationships between the data and parameters that may not be captured in an approximate theoretical model. Finally, the generated data are subdivided into training, validation, and test sets according to a 70/10/20\% split. The data generation, model implementation, and analysis codes are provided in the linked GitHub repository \cite{ml-pnr}. \\

\section*{Results}

\begin{figure*}[t]
    \centering
    \includegraphics[width=0.9\linewidth]{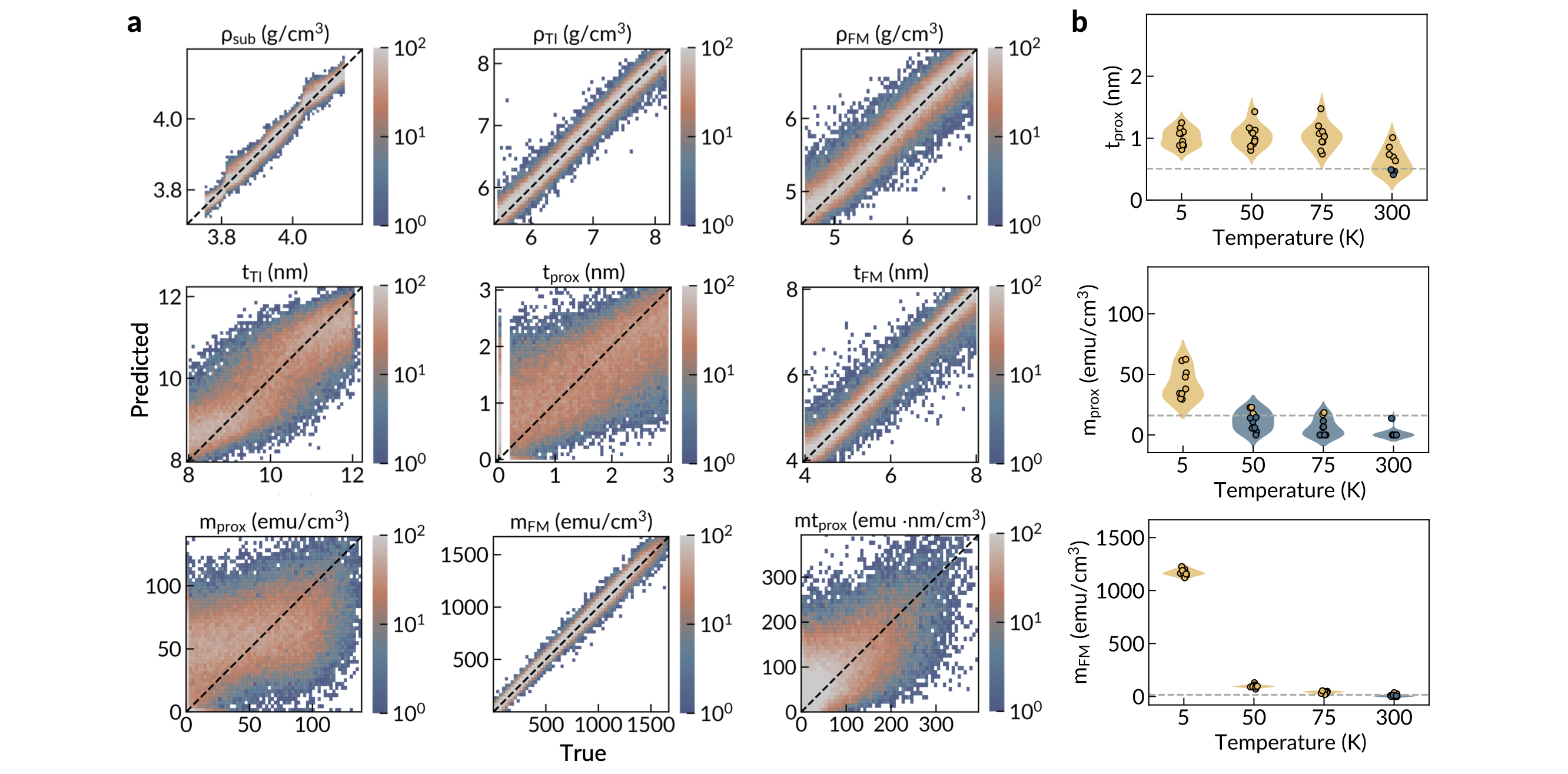}
    \caption{\textbf{Regressor performance and predictions on experimental data.} \textbf{a.} Histograms of predicted versus true values for different sample parameters of the test dataset. \textbf{b.} The predictions of proximity layer thickness t$_\text{prox}$ and magnetization m$_\text{prox}$, and FM layer magnetization m$_\text{FM}$ obtained from 10 instances of the VAE trained with different initial weights, shown as a function of the measurement temperature of the corresponding experiment. Gray dashed lines indicate the optimal thresholds obtained for proximity classification. Scattered points above (below) the determined threshold are colored yellow (blue). Violin plots of the predicted values for experiments with majority predictions above (below) the threshold are shaded yellow (blue).}
    \label{fig:Fig3}
\end{figure*}

We evaluate the trained VAE on its ability to recover the sample parameters from the experimental PNR profiles of the Bi$_2$Se$_3$/EuS system. The loss trajectories of the training and validation sets are shown in Fig. \ref{fig:FigS8}. We first compare the measured and decoded reflectometry profiles corresponding to four PNR experiments taken at different temperatures between 5 K and 300 K in Fig. \ref{fig:Fig2}a. These experimental PNR data are reproduced from Ref. \citenum{katmis2016high}. The left panel of Fig. \ref{fig:Fig2}a shows the reconstructed reflectometry profiles for both spin channels for the measurements at 5 K. The right panel shows the spin asymmetry ($R^{++} -R^{--})/(R^{++} +R^{--}$) calculated for both the measured and decoded profiles of the four experimental reflectivities at 5, 50, 75 and 300 K. Representative reconstructions of the test dataset in each error quartile are also shown in Fig. \ref{fig:FigS10}a. The four decoded experimental profiles are all found to be inliers of the distribution of reconstruction errors (Fig. \ref{fig:FigS10}b). This suggests that the chosen parameter ranges are likely suitable for the data under consideration. Next, using the parameter values predicted by the regressor, we calculate the SLD profiles for the measurements at each temperature (Fig. \ref{fig:Fig2}b). Note that the SLD profiles are computed directly using predicted parameter values and are not derived from the reconstructed PNR profiles. The nuclear (NSLD) and absorption (ASLD) scattering length density profiles appear largely consistent for the measurements at different temperatures, suggesting that the predicted values of the temperature-independent parameters, such as the thickness and density of each layer, are physically plausible. However, we do observe {\color{revision}a change in} the NSLD of the bulk FM layer {\color{revision}at 300K} that is worth mentioning. By examining the underlying parameters generating each SLD profile, we identify that the bulk {\color{revision}FM thickness increases slightly with temperature, which appears to coincide with slight reductions in the thickness of the TI and proximity layers, and a more significant reduction in the interface roughness.} At this stage, the exact origin of this temperature dependence is not well understood; the roughness values predicted for the FM and capping layers do not appear to follow a clear temperature-dependent trend and are likely prone to bigger uncertainties than the bulk parameters, but a possibility is that higher temperatures contribute to {\color{revision}smoothing the buried interfaces, such as those between the TI and FM, and FM and capping layers, which could partially explain the NSLD fluctuation.} The magnetic scattering length density (MSLD) profile is maximal at the EuS layer and exhibits a slight shoulder near the TI interface at 5 K, corresponding to the proximity layer. The MSLD magnitude drops progressively as the temperature is increased and disappears at 300K. These observations can be further traced back to the latent representations of the four experimental examples. In Fig. \ref{fig:Fig2}c, we visualize the latent space by projecting the encoded test dataset along the two dimensions with the largest local gradients for a given parameter value, e.g. substrate density $\rho_\text{sub}$. Specifically, the horizontal and vertical axes of each subplot correspond to the latent dimensions with the largest and second largest gradient of the target parameter, respectively. The local gradients $\partial v_{i}/\partial z_{j}$, where $v_{i}$ denotes the $i$-th parameter and $z_{j}$ denotes the $j$-th latent channel, are estimated using the 32 nearest-neighbors of each scattered point. Visualizations of all predicted parameters are given in Fig. \ref{fig:FigS12}. The scattered points, each corresponding to one profile of the test dataset, are colored according to the true value of the parameter viewed in each subplot. We find that the latent space is well-organized according to these parameter values, including the thickness t$_\text{prox}$ and magnetization m$_\text{prox}$ of the proximity layer. The latent representations of the four experimental PNR profiles are indicated by the outlined circles in the projection plots and are colored by the corresponding temperature. Notably, for temperature-independent quantities such as the TI density $\rho_\text{TI}$, the experimental points at different temperatures are generally insensitive to the gradient direction of the underlying parameter value, while for those like {\color{revision}the EuS magnetization m$_\text{FM}$} and m$_\text{prox}$, the points at different temperatures follow the gradient direction of the parameter values closely. This corroborates our observations that the trained VAE learns a sensible and interpretable latent representation of PNR profiles from which the physical parameter values may be estimated. The degree of parameter entanglement can be inferred from Fig. \ref{fig:Fig2}d, which shows the magnitudes of the average gradients of each sample parameter with respect to each latent channel. The {\color{revision}first 13} latent dimensions are conditioned to vary with the values of the corresponding parameter, while {\color{revision}the remaining channels are not explicitly linked to one specific} physical parameter, \textit{i.e.} they are regularized by the conventional standard normal prior distribution. {\color{revision}However, we observe that these ``free" channels sometimes participate in relating two or more parameters to one another, as observed for t$_\text{prox}$, t$_\text{TI}$, and t$_\text{FM}$ in Fig. \ref{fig:Fig2}d.}

{\color{revision}Next,} we assess the overall regression accuracy of our trained model on the test dataset for each of the sample parameters visualized in Fig. \ref{fig:Fig2}c. In each subplot of Fig. \ref{fig:Fig3}a, the test data points are histogrammed according to the true and predicted values of a given sample parameter. Corresponding plots for the complete set of predicted parameters are shown in Fig. \ref{fig:FigS14}a. {\color{revision}We also include the histogram for the parameter mt$_\text{prox}$, defined as the product of proximity layer thickness and magnetization, in the last panel of Fig. \ref{fig:Fig3}a.} The values of the bulk layer properties appear very well reproduced by the regressor, while t$_\text{prox}$ and m$_\text{prox}$, which exhibit much weaker signatures and tend to be expressed most in the noisier, high-$Q$ region of the PNR profiles, are somewhat underestimated {\color{revision}at large values and overestimated for non-proximity-coupled samples}. Note that the sharp discontinuity in the t$_\text{prox}$ histogram corresponds to the resolution threshold of {\color{revision}2 \AA}\ imposed on the generated data. To assess the reproducibility of the regression results, we trained ten identical models with different initial weights and collected statistics of the resulting predictions. Fig. \ref{fig:Fig3}b shows the predictions of these models for the values of t$_\text{prox}$, m$_\text{prox}$, and m$_\text{FM}$. We can optimize the trade-off between the true (tpr) and false (fpr) positive rates of correctly classifying proximity magnetism to obtain classification thresholds of thickness and magnetization that best separate the data points between the two classes. {\color{revision}These allow us to estimate the resolution threshold of the trained model to correctly distinguish samples with and without proximity magnetism within a certain confidence interval. The method of threshold determination is described in detail in the Supplementary Information and yields the} average classification thresholds of thickness and magnetization across the ten models indicated by the dashed gray lines in Fig. \ref{fig:Fig3}b. {\color{revision}The thresholds are found at 5.1 \AA\ and 16 emu cm$^{-3}$ (1 emu = 1 A$\cdot$m$^2$), corresponding to recalls of 85\% for $t_\text{prox}$, and 80\% for $m_\text{prox}$, for both the positive and negative classes (Fig. \ref{fig:FigS14}b-c)}. The optimal {\color{revision}thickness threshold is slightly higher than the resolution threshold of the generated data but corresponds well to the neutron wavelength of the reflectivity simulation, 4.75 \AA.} While a small spread in the predicted values across the ten models is observed, the overall trend in the predictions is as expected. Notably, all ten models predict t$_\text{prox}$ and m$_\text{prox}$ values above their respective thresholds at 5\ K. {\color{revision}The values of m$_\text{prox}$ decay with increasing temperature while t$_\text{prox}$ remains relatively constant until a slight drop at 300K.} Similarly, m$_\text{FM}$ drops rapidly beyond its Curie temperature. We note that the predicted values of m$_\text{prox}$ at intermediate temperatures are still often nonzero. This may be attributable to strong magnetic fluctuations above the EuS Curie temperature stabilizing a weak proximity effect below the resolution threshold of our model \cite{nogueira2012fluctuation,huang2020emergent}. Note that if weak proximity magnetism persists at high temperatures, it must be below the resolution threshold of our current model. A tailored network trained on a narrow range of parameters can potentially be devised to clarify even weaker signatures of proximity magnetism that may be expected at higher temperatures; however, the current model is highly suitable for surveying the evolution of proximity magnetism over a broad experimental parameter space, such as a wide temperature range. The predicted values obtained across the ten models for the remaining sample parameters, which are expected to be temperature-independent, are plotted in Fig. \ref{fig:FigS16}. Here, we observe similar consistency among most parameters, with the exception of fluctuations in the {\color{revision}roughnesses and FM layer thickness} noted in the previous discussion.

\subsection*{Resolving interfacial antiferromagnetic coupling}

\begin{figure*}[t]
    \centering
    \includegraphics[width=0.9\linewidth]{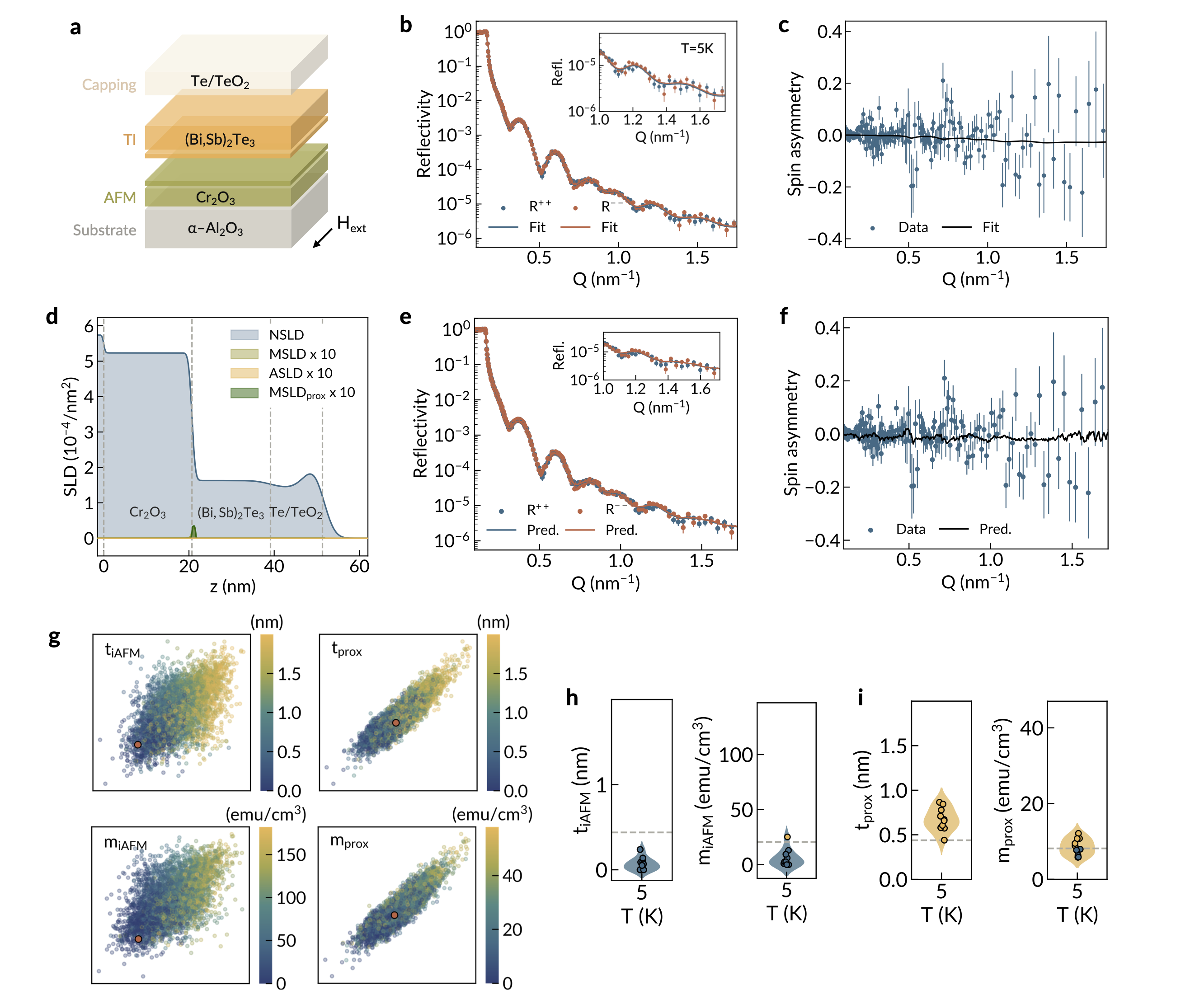}
    \caption{\textbf{Proximity magnetism in the TI-AFM system.} \textbf{a.} Schematic illustration of the (Bi,Sb)$_2$Te$_3$/Cr$_2$O$_3$ system. \textbf{b.} Experimental PNR profile at 5 K and best fit obtained with the GenX parameter refinement program. An expanded view of the splitting between the two channels at high $Q$ is shown in the inset. Error-bars represent ±1 standard deviation. \textbf{c.} The spin asymmetry ($R^{++} -R^{--})/(R^{++} +R^{--}$) calculated from the measured (points) and best fit (solid line) $R^{++}$ and $R^{--}$ profiles shown in \textbf{b}. \textbf{d.} {\color{revision}Nuclear (NSLD), magnetic (MSLD), and absorption (ASLD) scattering length density profiles obtained from the regressor predictions for the experimental measurement shown in \textbf{b}. The MSLD contribution from the proximity layer is shown in dark green.} \textbf{e.} Decoded profiles of the PNR measurement in \textbf{b}. An expanded view of the profiles at high $Q$ is shown in the inset. {\color{revision}\textbf{f.} Spin asymmetry calculated from the measured (points) and predicted (solid line) $R^{++}$ and $R^{--}$ profiles shown in \textbf{b}.} \textbf{g.} Projections of the latent encoding of the test dataset along the latent dimensions with the largest gradients for the thickness and magnetization of the interfacial AFM and TI layers. Red points show the latent encoding of the PNR profile in \textbf{b}. \textbf{h-i.} The predictions of \textbf{h.} interfacial AFM layer thickness and magnetization, and \textbf{i.} proximity layer thickness and magnetization obtained from 10 instances of the VAE trained with different initial weights. Gray dashed lines indicate the optimal thresholds for proximity classification. Scattered points above (below) the threshold are colored yellow (blue).}
    \label{fig:Fig4}
\end{figure*}

Lastly, we apply our approach to elucidate proximity magnetism from a more challenging PNR profile of an intrinsic TI (Bi,Sb)$_2$Te$_3$ interfaced with the AFM Cr$_2$O$_3$, shown schematically in Fig. \ref{fig:Fig4}a. {\color{revision} Bulk Cr$_2$O$_3$ is a well-known antiferromagnetic insulator with a N\'{e}el temperature of 307 K.} At the interface between a TI and AFM, magnetic atoms on the AFM surface have been shown to induce interfacial ferromagnetic order in the TI which can survive at much higher temperatures than that produced by doping or interfacing with a FM film, owing to the typically higher N\'{e}el temperatures \cite{he2017tailoring,he2018exchange,he2018topological,luo2013massive,wang2019observation,pan2020observation}. However, magnetic proximity coupling between an AFM and TI is comparatively weaker and thereby more challenging to isolate experimentally. Fig. \ref{fig:Fig4}b shows the experimental PNR profile of the intrinsic (Bi,Sb)$_2$Te$_3$/Cr$_2$O$_3$ system measured at the Magnetism Reflectometer at the Spallation Neutron Source (SNS) at Oak Ridge National Laboratory \cite{lauter2009highlights}. Data were collected at a temperature of 5 K with an in-plane external magnetic field of 1 T. The TI has a nominal composition of (Bi$_{0.2}$Sb$_{0.8}$)$_2$Te$_3$. {\color{revision} Additional experimental details are described in the Methods section.} Subtle evidence of spin splitting is observed in the associated plot of the spin asymmetry, ($R^{++} -R^{--})/(R^{++} +R^{--}$), shown in Fig. \ref{fig:Fig4}c. The experimental data in Figs. \ref{fig:Fig4}b-c are superimposed with the corresponding best fit obtained using the GenX parameter refinement program \cite{bjorck2007genx}. However, a major challenge encountered during conventional fitting is that repeated refinement with different initial populations fails to reproducibly predict the proximity effect, as the weak spin splitting observed in the PNR profiles could be attributed to either a small net magnetization at the AFM surface, or to proximity magnetism in the interfacial TI layer. To address this challenge, we train our VAE on a set of synthetic PNR profiles of the heterostructure shown in Fig. \ref{fig:Fig4}a. Similar to the case of Bi$_2$Se$_3$/EuS, we simulate the PNR profiles of {\color{revision}$200,000$} candidate systems parameterized by the density, thickness, and roughness of each layer, which included the TI, AFM, sapphire substrate, Te capping layer, and a possible TeO$_2$ surface film. In a subset of these examples, we model the presence of an interfacial FM layer on either the AFM or TI surface, or both, parameterized by thickness, roughness, and magnetization, and sharing the density value of the corresponding bulk layer. {\color{revision}Note that the predefined range for the magnetization of the interfacial layers is equivalent in terms of formula units but inequivalent in units of emu cm$^{-3}$ due to the very different densities of the TI and AFM materials (Fig. \ref{fig:FigS5}). The theoretical thickness resolution is likewise set to 2 \AA\ for both possible interfacial layers}. PNR profiles are simulated over the experimentally-measured $Q$-range from 0.1 to 1.72 nm$^{-1}$ and normalized to a maximum value of 1. The instrument background is sampled uniformly at random between {\color{revision}10$^{-8}$} and 10$^{-4}$ on a logarithmic scale. The remaining data preparation steps are conducted as for the Bi$_2$Se$_3$/EuS example. The reconstructed PNR profile, shown in Fig. \ref{fig:Fig4}e, matches the experimental data closely and is in the top reconstruction error quartile (Fig. \ref{fig:FigS11}). The corresponding SLD profile obtained using the predicted sample parameters is shown in Fig. \ref{fig:Fig4}d, where an interfacial FM layer is evidenced by the peak in the MSLD profile at the TI surface. {\color{revision}We also plot the spin asymmetry in Fig. \ref{fig:Fig4}f, which displays a weak non-zero signature near 0.5 nm$^{-1}$. Most significantly,} the trained VAE yields a latent representation for the test dataset and experimental example as shown in Fig. \ref{fig:Fig4}g. The points in each subplot are colored according to the true values of the thickness or magnetization of the interfacial FM layer on either the AFM surface, denoted t$_\text{iAFM}$ and m$_\text{iAFM}$, or the TI surface, denoted t$_\text{prox}$ and m$_\text{prox}$, respectively. The remaining sample parameters predicted by the model are plotted in Fig. \ref{fig:FigS13}a. We see that the experimental PNR profile is mapped unambiguously to a region with t$_\text{iAFM} \approx 0$ and m$_\text{iAFM} \approx 0$, while t$_\text{prox}$ and m$_\text{prox}$ are predicted to be approximately {\color{revision}8.4 \AA\ and 12 emu cm$^{-3}$}, respectively. The regression accuracy for each predicted parameter is plotted in Fig. \ref{fig:FigS15}a. To test the robustness of these predictions, we train another ten identical models with different initial weights. Figs. \ref{fig:Fig4}h and i show the predictions of the ten models for the values of t$_\text{iAFM}$ and m$_\text{iAFM}$, and t$_\text{prox}$ and m$_\text{prox}$, respectively. The gray dashed lines delimiting proximity-coupled examples are similarly obtained by optimizing the trade-off between the true and false positive rates of correctly classifying proximity magnetism in the validation set for each model. The average threshold values for the proximity layer are found at {\color{revision}4.4 \AA\ and 8.2 emu cm$^{-3}$}, corresponding to {\color{revision}recalls of 76 and 72\%} in both classes, respectively (Fig. \ref{fig:FigS15}b-c). {\color{revision}Since the values of m$_\text{iAFM}$ are roughly three times as large as those of m$_\text{prox}$ in conventional units, the resolution threshold for m$_\text{iAFM}$ is computed separately following the same procedure and found to be 20.6 emu cm$^{-3}$. Using these thresholds, we find that only half of the models predict proximity magnetism is present when considering the predicted m$_\text{prox}$ values, though t$_\text{prox}$ is well-resolved by most models; however, almost all models unambiguously predict t$_\text{iAFM}$ and m$_\text{iAFM}$ well below their respective thresholds.} Thus, although the VAE approach pushes the boundary for resolving subtle magnetic signatures, it is possible that a very weak proximity effect is present at or slightly below the resolution threshold we could achieve with the current model. Nonetheless, the predictions could be used as a valuable screening tool before conducting finer measurements with either longer acquisition times or at higher $Q$ to more clearly resolve the spin splitting, which could potentially benefit experimental planning and optimize the use of scientific user facilities.

\section*{Discussion}
Machine learning methods are valuable means of uncovering hidden patterns in materials data and elucidating the relationships between structural descriptors and measured quantities. However, it is often desirable to balance the flexibility of ``black box" neural networks with a degree of interpretability in terms of physical parameters. We accomplish this in our VAE-based framework by conditioning the latent channels to emulate the behavior of the original sample parameters, which enables direct, visual inspection of encoded profiles in terms of meaningful physical quantities. However, our assessment of parameter entanglement in Fig. \ref{fig:Fig2}d reveals underlying correlations between sample parameters; for example, proximity layer thickness and magnetization are deeply entwined, since finite proximity magnetization implies finite thickness of the interfacial layer, and vice versa. This suggests that the prior assumption that all latent dimensions are sampled from independent normal distributions does not perfectly describe a latent space that is conditioned to vary directly with certain correlated physical parameters. A possible improvement to the existing approach would be to describe the latent space in terms of several joint distributions of a few strongly-correlated parameters, which can be tuned to balance the number of necessary network parameters. We note on a few additional considerations for future work. In particular, density fluctuations may be present in certain samples and can require fitting a number of distinct sub-layers of each material. {\color{revision}Additionally, a more comprehensive study of the effects of noise on the VAE outcomes would be relevant to determine the effectiveness of such a model to screen candidate systems.} These and other specialized features can be readily integrated into the framework presented in this work. {\color{revision}Finally, it is important to acknowledge that while the present work aims to reduce reliance on expert insight for PNR parameter retrieval, domain knowledge is still needed to construct a layered representation of the system that considers all relevant features for generating the training data, as well as in the interpretation of the final results. Machine learning-driven discovery in this domain might entail more sophisticated models that not only determine the suitability of a particular hypothesis, such as the presence or absence of proximity magnetism, but rather discover the plausible mechanisms underlying a given observation.}

\section*{Conclusion}
A quantitative understanding of structural and magnetic information encoded in PNR measurements is often critical to resolving important interfacial phenomena, but experimental factors and lack of adequate fitting constraints can impede parameter retrieval without expert insight. In this work, we construct a data-driven framework for PNR parameter retrieval by training a conditioned VAE to map reflectometry profiles with moderate experimental noise to a well-organized, low-dimensional space from which sample parameters can be readily obtained. We balance the flexibility and interpretability of our model through latent space engineering, enabling in-depth analysis of the resulting predictions. Compared to traditional fitting methods, our framework involves minimal user intervention overall, requiring no expert insight for parameter initialization or refinement, yet is capable of resolving parameter values near the experimental resolution limit. It further enables evaluation of the entire parameter search space by readily identifying outliers of the chosen domain. A possible extension of the framework is suggested to account for intrinsic correlations between conditioning variables. We apply our method to recover the SLD profiles of two proximity-coupled systems at subnanometer resolution, and we envision its potential application to a broader context of elusive phases expressed through weak experimental signatures, such as the axion insulator and topological superconducting phases. We anticipate that the methodology developed in this work can facilitate the development of comprehensive and fully-automated analysis routines for PNR parameter retrieval of a broad range of materials systems, as well as inform the wide spectrum of spectroscopic analysis workflows requiring parameter refinement.

{\color{revision} \section*{Methods}}
{\color{revision} \subsection*{Training details}}
{\color{revision} Training data were generated using a Python implementation of the GenX neutron reflectivity modeling code \cite{bjorck2007genx}. Simulation of the $200,000$ PNR profiles took approximately 48 sec. using 25 parallel processes on Intel(R) Xeon(R) Gold 5218 processors. Neural network models were implemented in Python using the PyTorch \cite{NEURIPS2019_9015} libraries and trained on a Quadro RTX 6000 graphics processing unit (GPU) with $24\,\text{GB}$ of random access memory (RAM). For the architecture used in this work, training a single epoch took $\sim$45 sec., and each model was trained over 100 epochs. Additional details of the network architecture and final hyperparameters are provided in the Supplementary Information.}

{\color{revision} \subsection*{Experimental details}}
{\color{revision} Intrinsic (Bi,Sb)$_2$Te$_3$ consisting of a $\sim$15 quintuple-layer (QL) of (Bi,Sb)$_2$Te$_3$ with nominal composition (Bi$_{0.2}$Sb$_{0.8}$)$_2$Te$_3$ was grown by molecular beam epitaxy (MBE) on a thin film of the AFM insulator Cr$_2$O$_3$ with a nominal thickness of 20 nm. The Bi:Sb ratio was optimized to locate the Fermi level near the Dirac node of the electronic surface states. The AFM Cr$_2$O$_3$ film was grown on a sapphire substrate in a pulsed laser deposition chamber with a base pressure of 2$\times$10$^{-8}$ mbar. A $\sim$10 nm amorphous Te capping layer was deposited on top of the (Bi,Sb)$_2$Te$_3$ film to protect from degradation. PNR experiments were carried out at the Magnetism Reflectometer at the Spallation Neutron Source (SNS) at Oak Ridge National Laboratory at a temperature of 5 K with a 1 T in-plane magnetic field.}
\\

{\color{revision} \section*{Supplementary Material}}
{\color{revision} See supplemental material for additional details regarding data preparation and noise estimation; the VAE architecture and training history; the complete set of latent space visualizations; and the full set of regressor performance results, including the method of threshold determination.}

\begin{acknowledgments}
\noindent N.A., Z.C., and M.L. thank C.H. Rycroft for helpful discussions. N.A., Z.C., and M.L. acknowledge the support from U.S. DOE, Office of Science (SC), Basic Energy Sciences (BES), awards No. DE-SC0020148 and DE-SC0021940. N.A. acknowledges the support of the National Science Foundation Graduate Research Fellowship Program under Grant No. 1122374. M.L. is partially supported by DOE DE-AR0001298, NSF DMR-2118448, and Norman C. Rasmussen Career Development Chair. This research used resources at the Spallation Neutron Source, a DOE Office of Science User Facility operated by the Oak Ridge National Laboratory.
\end{acknowledgments}

\bibliography{main}

\begin{thebibliography}{10}

\bibitem{fitzsimmons2004neutron}
M.~R. Fitzsimmons, S.~Bader, J.~Borchers, G.~Felcher, J.~Furdyna, A.~Hoffmann,
  J.~Kortright, I.~K. Schuller, T.~Schulthess, S.~Sinha, {\em et~al.},
  ``Neutron scattering studies of nanomagnetism and artificially structured
  materials,'' {\em Journal of Magnetism and Magnetic Materials}, vol.~271,
  no.~1, pp.~103--146, 2004.

\bibitem{lauter2007neutron}
V.~Lauter-Pasyuk, ``Neutron grazing incidence techniques for nano-science,''
  {\em Collection SFN}, vol.~7, pp.~s221--s240, 2007.

\bibitem{fitzsimmons2007pinned}
M.~Fitzsimmons, B.~Kirby, S.~Roy, Z.-P. Li, I.~V. Roshchin, S.~Sinha, and I.~K.
  Schuller, ``Pinned magnetization in the antiferromagnet and ferromagnet of an
  exchange bias system,'' {\em Physical Review B}, vol.~75, no.~21, p.~214412,
  2007.

\bibitem{bennett2016giant}
S.~Bennett, A.~Wong, A.~Glavic, A.~Herklotz, C.~Urban, I.~Valmianski,
  M.~Biegalski, H.~Christen, T.~Ward, and V.~Lauter, ``Giant controllable
  magnetization changes induced by structural phase transitions in a
  metamagnetic artificial multiferroic,'' {\em Scientific Reports}, vol.~6,
  no.~1, pp.~1--7, 2016.

\bibitem{gilbert2016structural}
D.~A. Gilbert, A.~J. Grutter, E.~Arenholz, K.~Liu, B.~J. Kirby, J.~A. Borchers,
  and B.~B. Maranville, ``Structural and magnetic depth profiles of
  magneto-ionic heterostructures beyond the interface limit,'' {\em Nature
  Communications}, vol.~7, no.~1, pp.~1--8, 2016.

\bibitem{theis2020self}
K.~Theis-Br{\"o}hl, A.~Saini, M.~Wolff, J.~A. Dura, B.~B. Maranville, and J.~A.
  Borchers, ``Self-assembly of magnetic nanoparticles in ferrofluids on
  different templates investigated by neutron reflectometry,'' {\em
  Nanomaterials}, vol.~10, no.~6, p.~1231, 2020.

\bibitem{need2020magnetic}
R.~F. Need, S.-K. Bac, X.~Liu, S.~Lee, B.~J. Kirby, M.~Dobrowolska, J.~Kossut,
  and J.~K. Furdyna, ``Magnetic properties and electronic origin of the
  interface between dilute magnetic semiconductors with orthogonal magnetic
  anisotropy,'' {\em Physical Review Materials}, vol.~4, no.~5, p.~054410,
  2020.

\bibitem{keunecke2020high}
M.~Keunecke, F.~Lyzwa, D.~Schwarzbach, V.~Roddatis, N.~Gauquelin,
  K.~M{\"u}ller-Caspary, J.~Verbeeck, S.~J. Callori, F.~Klose, M.~Jungbauer,
  {\em et~al.}, ``High-{T}$_c$ interfacial ferromagnetism in
  {S}r{M}n{O}$_3$/{L}a{M}n{O}$_3$ superlattices,'' {\em Advanced Functional
  Materials}, vol.~30, no.~18, p.~1808270, 2020.

\bibitem{liu2021ferroelectric}
C.~Liu, Y.~Liu, B.~Zhang, C.-J. Sun, D.~Lan, P.~Chen, X.~Wu, P.~Yang, X.~Yu,
  T.~Charlton, {\em et~al.}, ``Ferroelectric self-polarization controlled
  magnetic stratification and magnetic coupling in ultrathin
  {L}a$_{0.67}${S}r$_{0.33}${M}n{O}$_3$ films,'' {\em ACS Applied Materials \&
  Interfaces}, 2021.

\bibitem{bhatnagar2021differentiation}
T.~Bhatnagar-Sch{\"o}ffmann, E.~Kentzinger, A.~Sarkar, P.~Sch{\"o}ffmann,
  Q.~Lan, L.~Jin, A.~Kovacs, A.~Grutter, B.~Kirby, R.~Beerwerth, {\em et~al.},
  ``Differentiation between strain and charge mediated magnetoelectric coupling
  in
  {L}a$_{0.7}${S}r$_{0.3}${M}n{O}$_3$/{P}b({M}g$_{1/3}${N}b$_{2/3}$)$_{0.7}${T}i$_{0.3}${O}$_3$
  (001),'' {\em New Journal of Physics}, 2021.

\bibitem{wang2021optically}
M.~Wang, H.~Xu, T.~Wu, H.~Ambaye, J.~Qin, J.~Keum, I.~N. Ivanov, V.~Lauter, and
  B.~Hu, ``Optically induced static magnetization in metal halide perovskite
  for spin-related optoelectronics,'' {\em Advanced Science}, vol.~8, no.~11,
  p.~2004488, 2021.

\bibitem{ankner1999polarized}
J.~Ankner and G.~Felcher, ``Polarized-neutron reflectometry,'' {\em Journal of
  Magnetism and Magnetic Materials}, vol.~200, no.~1-3, pp.~741--754, 1999.

\bibitem{majkrzak2006polarized}
C.~Majkrzak, K.~O'Donovan, and N.~Berk, ``Polarized neutron reflectometry,'' in
  {\em Neutron Scattering from Magnetic Materials}, pp.~397--471, Elsevier,
  2006.

\bibitem{toperverg2015polarized}
B.~P. Toperverg, ``Polarized neutron reflectometry of magnetic
  nanostructures,'' {\em The Physics of Metals and Metallography}, vol.~116,
  no.~13, pp.~1337--1375, 2015.

\bibitem{nichols2016emerging}
J.~Nichols, X.~Gao, S.~Lee, T.~L. Meyer, J.~W. Freeland, V.~Lauter, D.~Yi,
  J.~Liu, D.~Haskel, J.~R. Petrie, {\em et~al.}, ``Emerging magnetism and
  anomalous hall effect in iridate--manganite heterostructures,'' {\em Nature
  Communications}, vol.~7, no.~1, pp.~1--6, 2016.

\bibitem{zhan2019probing}
X.~Zhan, G.~Li, J.~Cai, T.~Zhu, J.~Cooper, C.~Kinane, and S.~Langridge,
  ``Probing the transfer of the exchange bias effect by polarized neutron
  reflectometry,'' {\em Scientific Reports}, vol.~9, no.~1, pp.~1--9, 2019.

\bibitem{bhattacharyya2021recent}
S.~Bhattacharyya, G.~Akhgar, M.~Gebert, J.~Karel, M.~T. Edmonds, and M.~S.
  Fuhrer, ``Recent progress in proximity coupling of magnetism to topological
  insulators,'' {\em Advanced Materials}, p.~2007795, 2021.

\bibitem{vobornik2011magnetic}
I.~Vobornik, U.~Manju, J.~Fujii, F.~Borgatti, P.~Torelli, D.~Krizmancic, Y.~S.
  Hor, R.~J. Cava, and G.~Panaccione, ``Magnetic proximity effect as a pathway
  to spintronic applications of topological insulators,'' {\em Nano letters},
  vol.~11, no.~10, pp.~4079--4082, 2011.

\bibitem{eremeev2013magnetic}
S.~Eremeev, V.~Men'Shov, V.~Tugushev, P.~M. Echenique, and E.~V. Chulkov,
  ``Magnetic proximity effect at the three-dimensional topological
  insulator/magnetic insulator interface,'' {\em Physical Review B}, vol.~88,
  no.~14, p.~144430, 2013.

\bibitem{lang2014proximity}
M.~Lang, M.~Montazeri, M.~C. Onbasli, X.~Kou, Y.~Fan, P.~Upadhyaya, K.~Yao,
  F.~Liu, Y.~Jiang, W.~Jiang, {\em et~al.}, ``Proximity induced
  high-temperature magnetic order in topological insulator-ferrimagnetic
  insulator heterostructure,'' {\em Nano letters}, vol.~14, no.~6,
  pp.~3459--3465, 2014.

\bibitem{lee2014magnetic}
A.~T. Lee, M.~J. Han, and K.~Park, ``Magnetic proximity effect and spin-orbital
  texture at the {B}i$_2$ {S}e$_3$/{E}u{S} interface,'' {\em Physical Review
  B}, vol.~90, no.~15, p.~155103, 2014.

\bibitem{li2015magnetic}
M.~Li, W.~Cui, J.~Yu, Z.~Dai, Z.~Wang, F.~Katmis, W.~Guo, and J.~Moodera,
  ``Magnetic proximity effect and interlayer exchange coupling of
  ferromagnetic/topological insulator/ferromagnetic trilayer,'' {\em Physical
  Review B}, vol.~91, no.~1, p.~014427, 2015.

\bibitem{liu2015enhancing}
W.~Liu, L.~He, Y.~Xu, K.~Murata, M.~C. Onbasli, M.~Lang, N.~J. Maltby, S.~Li,
  X.~Wang, C.~A. Ross, {\em et~al.}, ``Enhancing magnetic ordering in cr-doped
  {B}i$_2${S}e$_3$ using high-t$_c$ ferrimagnetic insulator,'' {\em Nano
  Letters}, vol.~15, no.~1, pp.~764--769, 2015.

\bibitem{katmis2016high}
F.~Katmis, V.~Lauter, F.~S. Nogueira, B.~A. Assaf, M.~E. Jamer, P.~Wei,
  B.~Satpati, J.~W. Freeland, I.~Eremin, D.~Heiman, {\em et~al.}, ``A
  high-temperature ferromagnetic topological insulating phase by proximity
  coupling,'' {\em Nature}, vol.~533, no.~7604, pp.~513--516, 2016.

\bibitem{he2017tailoring}
Q.~L. He, X.~Kou, A.~J. Grutter, G.~Yin, L.~Pan, X.~Che, Y.~Liu, T.~Nie,
  B.~Zhang, S.~M. Disseler, {\em et~al.}, ``Tailoring exchange couplings in
  magnetic topological-insulator/antiferromagnet heterostructures,'' {\em
  Nature Materials}, vol.~16, no.~1, pp.~94--100, 2017.

\bibitem{che2018proximity}
X.~Che, K.~Murata, L.~Pan, Q.~L. He, G.~Yu, Q.~Shao, G.~Yin, P.~Deng, Y.~Fan,
  B.~Ma, {\em et~al.}, ``Proximity-induced magnetic order in a transferred
  topological insulator thin film on a magnetic insulator,'' {\em ACS Nano},
  vol.~12, no.~5, pp.~5042--5050, 2018.

\bibitem{koren2018magnetic}
G.~Koren, ``Magnetic proximity effect of a topological insulator and a
  ferromagnet in thin-film bilayers of {B}i$_{0.5}${S}b$_{1.5}${T}e$_3$ and
  {S}r{R}u{O}$_3$,'' {\em Physical Review B}, vol.~97, no.~5, p.~054405, 2018.

\bibitem{he2018exchange}
Q.~L. He, G.~Yin, A.~J. Grutter, L.~Pan, X.~Che, G.~Yu, D.~A. Gilbert, S.~M.
  Disseler, Y.~Liu, P.~Shafer, {\em et~al.}, ``Exchange-biasing topological
  charges by antiferromagnetism,'' {\em Nature Communications}, vol.~9, no.~1,
  pp.~1--8, 2018.

\bibitem{he2018topological}
Q.~L. He, G.~Yin, L.~Yu, A.~J. Grutter, L.~Pan, C.-Z. Chen, X.~Che, G.~Yu,
  B.~Zhang, Q.~Shao, {\em et~al.}, ``Topological transitions induced by
  antiferromagnetism in a thin-film topological insulator,'' {\em Physical
  Review Letters}, vol.~121, no.~9, p.~096802, 2018.

\bibitem{hou2019magnetizing}
Y.~Hou, J.~Kim, and R.~Wu, ``Magnetizing topological surface states of
  {B}i$_2${S}e$_3$ with a {C}r{I}$_3$ monolayer,'' {\em Science Advances},
  vol.~5, no.~5, p.~eaaw1874, 2019.

\bibitem{akiyama2019direct}
R.~Akiyama, R.~Ishikawa, K.~Akutsu, R.~Nakanishi, Y.~Tomohiro, K.~Watanabe,
  K.~Iida, M.~Mitome, S.~Hasegawa, and S.~Kuroda, ``Direct probe of
  ferromagnetic proximity effect at the interface in {F}e/{S}n{T}e
  heterostructure by polarized neutron reflectometry,'' {\em arXiv preprint
  arXiv:1910.10540}, 2019.

\bibitem{watanabe2019quantum}
R.~Watanabe, R.~Yoshimi, M.~Kawamura, M.~Mogi, A.~Tsukazaki, X.~Yu,
  K.~Nakajima, K.~S. Takahashi, M.~Kawasaki, and Y.~Tokura, ``Quantum anomalous
  hall effect driven by magnetic proximity coupling in all-telluride based
  heterostructure,'' {\em Applied Physics Letters}, vol.~115, no.~10,
  p.~102403, 2019.

\bibitem{pan2020observation}
L.~Pan, A.~Grutter, P.~Zhang, X.~Che, T.~Nozaki, A.~Stern, M.~Street, B.~Zhang,
  B.~Casas, Q.~L. He, {\em et~al.}, ``Observation of quantum anomalous hall
  effect and exchange interaction in topological insulator/antiferromagnet
  heterostructure,'' {\em Advanced Materials}, vol.~32, no.~34, p.~2001460,
  2020.

\bibitem{li2015proximity}
M.~Li, C.-Z. Chang, B.~J. Kirby, M.~E. Jamer, W.~Cui, L.~Wu, P.~Wei, Y.~Zhu,
  D.~Heiman, J.~Li, {\em et~al.}, ``Proximity-driven enhanced magnetic order at
  ferromagnetic-insulator--magnetic-topological-insulator interface,'' {\em
  Physical review letters}, vol.~115, no.~8, p.~087201, 2015.

\bibitem{tokura2019magnetic}
Y.~Tokura, K.~Yasuda, and A.~Tsukazaki, ``Magnetic topological insulators,''
  {\em Nature Reviews Physics}, vol.~1, no.~2, pp.~126--143, 2019.

\bibitem{yu2010quantized}
R.~Yu, W.~Zhang, H.-J. Zhang, S.-C. Zhang, X.~Dai, and Z.~Fang, ``Quantized
  anomalous hall effect in magnetic topological insulators,'' {\em Science},
  vol.~329, no.~5987, pp.~61--64, 2010.

\bibitem{kou2014scale}
X.~Kou, S.-T. Guo, Y.~Fan, L.~Pan, M.~Lang, Y.~Jiang, Q.~Shao, T.~Nie,
  K.~Murata, J.~Tang, {\em et~al.}, ``Scale-invariant quantum anomalous hall
  effect in magnetic topological insulators beyond the two-dimensional limit,''
  {\em Physical review letters}, vol.~113, no.~13, p.~137201, 2014.

\bibitem{kou2015magnetic}
X.~Kou, Y.~Fan, M.~Lang, P.~Upadhyaya, and K.~L. Wang, ``Magnetic topological
  insulators and quantum anomalous hall effect,'' {\em Solid State
  Communications}, vol.~215, pp.~34--53, 2015.

\bibitem{mogi2019large}
M.~Mogi, T.~Nakajima, V.~Ukleev, A.~Tsukazaki, R.~Yoshimi, M.~Kawamura, K.~S.
  Takahashi, T.~Hanashima, K.~Kakurai, T.-h. Arima, {\em et~al.}, ``Large
  anomalous hall effect in topological insulators with proximitized
  ferromagnetic insulators,'' {\em Physical Review Letters}, vol.~123, no.~1,
  p.~016804, 2019.

\bibitem{mogi2017magnetic}
M.~Mogi, M.~Kawamura, R.~Yoshimi, A.~Tsukazaki, Y.~Kozuka, N.~Shirakawa,
  K.~Takahashi, M.~Kawasaki, and Y.~Tokura, ``A magnetic heterostructure of
  topological insulators as a candidate for an axion insulator,'' {\em Nature
  Materials}, vol.~16, no.~5, pp.~516--521, 2017.

\bibitem{mogi2017tailoring}
M.~Mogi, M.~Kawamura, A.~Tsukazaki, R.~Yoshimi, K.~S. Takahashi, M.~Kawasaki,
  and Y.~Tokura, ``Tailoring tricolor structure of magnetic topological
  insulator for robust axion insulator,'' {\em Science Advances}, vol.~3,
  no.~10, p.~eaao1669, 2017.

\bibitem{xiao2018realization}
D.~Xiao, J.~Jiang, J.-H. Shin, W.~Wang, F.~Wang, Y.-F. Zhao, C.~Liu, W.~Wu,
  M.~H. Chan, N.~Samarth, {\em et~al.}, ``Realization of the axion insulator
  state in quantum anomalous hall sandwich heterostructures,'' {\em Physical
  Review Letters}, vol.~120, no.~5, p.~056801, 2018.

\bibitem{liang2017magnetic}
X.~Liang, L.~Deng, F.~Huang, T.~Tang, C.~Wang, Y.~Zhu, J.~Qin, Y.~Zhang,
  B.~Peng, and L.~Bi, ``The magnetic proximity effect and electrical field
  tunable valley degeneracy in {M}o{S}$_2$/{E}u{S} van der waals
  heterojunctions,'' {\em Nanoscale}, vol.~9, no.~27, pp.~9502--9509, 2017.

\bibitem{karpiak2019magnetic}
B.~Karpiak, A.~W. Cummings, K.~Zollner, M.~Vila, D.~Khokhriakov, A.~M. Hoque,
  A.~Dankert, P.~Svedlindh, J.~Fabian, S.~Roche, {\em et~al.}, ``Magnetic
  proximity in a van der waals heterostructure of magnetic insulator and
  graphene,'' {\em 2D Materials}, vol.~7, no.~1, p.~015026, 2019.

\bibitem{tong2019magnetic}
Q.~Tong, M.~Chen, and W.~Yao, ``Magnetic proximity effect in a van der waals
  moir{\'e} superlattice,'' {\em Physical Review Applied}, vol.~12, no.~2,
  p.~024031, 2019.

\bibitem{behera2019proximity}
S.~K. Behera, M.~Bora, S.~S.~P. Chowdhury, and P.~Deb, ``Proximity effects in
  graphene and ferromagnetic {C}r{B}r$_3$ van der waals heterostructures,''
  {\em Physical Chemistry Chemical Physics}, vol.~21, no.~46, pp.~25788--25796,
  2019.

\bibitem{huang2020emergent}
B.~Huang, M.~A. McGuire, A.~F. May, D.~Xiao, P.~Jarillo-Herrero, and X.~Xu,
  ``Emergent phenomena and proximity effects in two-dimensional magnets and
  heterostructures,'' {\em Nature Materials}, vol.~19, no.~12, pp.~1276--1289,
  2020.

\bibitem{zhao2020magnetic}
W.~Zhao, Z.~Fei, T.~Song, H.~K. Choi, T.~Palomaki, B.~Sun, P.~Malinowski, M.~A.
  McGuire, J.-H. Chu, X.~Xu, {\em et~al.}, ``Magnetic proximity and
  nonreciprocal current switching in a monolayer {WT}e$_2$ helical edge,'' {\em
  Nature Materials}, vol.~19, no.~5, pp.~503--507, 2020.

\bibitem{tang2020magnetic}
C.~Tang, Z.~Zhang, S.~Lai, Q.~Tan, and W.~Gao, ``Magnetic proximity effect in
  graphene/{C}r{B}r$_3$ van der waals heterostructures,'' {\em Advanced
  Materials}, vol.~32, no.~16, p.~1908498, 2020.

\bibitem{zhong2020layer}
D.~Zhong, K.~L. Seyler, X.~Linpeng, N.~P. Wilson, T.~Taniguchi, K.~Watanabe,
  M.~A. McGuire, K.-M.~C. Fu, D.~Xiao, W.~Yao, {\em et~al.}, ``Layer-resolved
  magnetic proximity effect in van der waals heterostructures,'' {\em Nature
  Nanotechnology}, vol.~15, no.~3, pp.~187--191, 2020.

\bibitem{dayen2020two}
J.-F. Dayen, S.~J. Ray, O.~Karis, I.~J. Vera-Marun, and M.~V. Kamalakar,
  ``Two-dimensional van der waals spinterfaces and magnetic-interfaces,'' {\em
  Applied Physics Reviews}, vol.~7, no.~1, p.~011303, 2020.

\bibitem{zhang2020controllable}
Y.~Zhang, K.~Shinokita, K.~Watanabe, T.~Taniguchi, M.~Goto, D.~Kan,
  Y.~Shimakawa, Y.~Moritomo, T.~Nishihara, Y.~Miyauchi, {\em et~al.},
  ``Controllable magnetic proximity effect and charge transfer in 2d
  semiconductor and double-layered perovskite manganese oxide van der waals
  heterostructure,'' {\em Advanced Materials}, vol.~32, no.~50, p.~2003501,
  2020.

\bibitem{zhang2020proximity}
L.~Zhang, X.~Huang, H.~Dai, M.~Wang, H.~Cheng, L.~Tong, Z.~Li, X.~Han, X.~Wang,
  L.~Ye, {\em et~al.}, ``Proximity-coupling-induced significant enhancement of
  coercive field and curie temperature in 2d van der waals heterostructures,''
  {\em Advanced Materials}, vol.~32, no.~38, p.~2002032, 2020.

\bibitem{bora2021magnetic}
M.~Bora and P.~Deb, ``Magnetic proximity effect in two-dimensional van der
  waals heterostructure,'' {\em Journal of Physics: Materials}, vol.~4, no.~3,
  p.~034014, 2021.

\bibitem{parratt1954surface}
L.~G. Parratt, ``Surface studies of solids by total reflection of x-rays,''
  {\em Physical Review}, vol.~95, no.~2, p.~359, 1954.

\bibitem{abeles1948propagation}
F.~Abel{\`e}s, ``Sur la propagation des ondes {\'e}lectromagn{\'e}tiques dans
  les milieux sratifi{\'e}s,'' in {\em Annales de Physique}, vol.~12,
  pp.~504--520, EDP Sciences, 1948.

\bibitem{sivia1991novel}
D.~Sivia, W.~Hamilton, G.~Smith, T.~Rieker, and R.~Pynn, ``A novel experimental
  procedure for removing ambiguity from the interpretation of neutron and x-ray
  reflectivity measurements:‘‘speckle holography’’,'' {\em Journal of
  applied physics}, vol.~70, no.~2, pp.~732--738, 1991.

\bibitem{de1995retrieval}
V.-O. De~Haan, A.~Van~Well, S.~Adenwalla, and G.~Felcher, ``Retrieval of phase
  information in neutron reflectometry,'' {\em Physical Review B}, vol.~52,
  no.~15, p.~10831, 1995.

\bibitem{pleshanov1999polarized}
N.~Pleshanov, ``Polarized neutron reflectometry with phase analysis,'' {\em
  Physica B: Condensed Matter}, vol.~269, no.~1, pp.~79--94, 1999.

\bibitem{o2002pinpointing}
K.~O'Donovan, J.~Borchers, C.~Majkrzak, O.~Hellwig, and E.~Fullerton,
  ``Pinpointing chiral structures with front-back polarized neutron
  reflectometry,'' {\em Physical review letters}, vol.~88, no.~6, p.~067201,
  2002.

\bibitem{durant2021optimising}
J.~H. Durant, L.~Wilkins, and J.~F. Cooper, ``Optimising experimental design in
  neutron reflectometry,'' {\em arXiv preprint arXiv:2108.05605}, 2021.

\bibitem{sivia1998bayesian}
D.~Sivia and J.~Webster, ``The bayesian approach to reflectivity data,'' {\em
  Physica B: Condensed Matter}, vol.~248, no.~1-4, pp.~327--337, 1998.

\bibitem{mccluskey2020general}
A.~R. McCluskey, J.~F. Cooper, T.~Arnold, and T.~Snow, ``A general approach to
  maximise information density in neutron reflectometry analysis,'' {\em
  Machine Learning: Science and Technology}, vol.~1, no.~3, p.~035002, 2020.

\bibitem{bjorck2007genx}
M.~Bj{\"o}rck and G.~Andersson, ``Genx: an extensible x-ray reflectivity
  refinement program utilizing differential evolution,'' {\em Journal of
  Applied Crystallography}, vol.~40, no.~6, pp.~1174--1178, 2007.

\bibitem{maranville2017interactive}
B.~B. Maranville, ``Interactive, web-based calculator of neutron and x-ray
  reflectivity,'' {\em J. Res. Natl. Inst. Stand. Technol}, vol.~122, no.~1,
  2017.

\bibitem{danauskas2008stochastic}
S.~M. Danauskas, D.~Li, M.~Meron, B.~Lin, and K.~Y.~C. Lee, ``Stochastic
  fitting of specular x-ray reflectivity data using stochfit,'' {\em Journal of
  Applied Crystallography}, vol.~41, no.~6, pp.~1187--1193, 2008.

\bibitem{greco2019fast}
A.~Greco, V.~Starostin, C.~Karapanagiotis, A.~Hinderhofer, A.~Gerlach,
  L.~Pithan, S.~Liehr, F.~Schreiber, and S.~Kowarik, ``Fast fitting of
  reflectivity data of growing thin films using neural networks,'' {\em Journal
  of applied crystallography}, vol.~52, no.~6, pp.~1342--1347, 2019.

\bibitem{mironov2021towards}
D.~Mironov, J.~H. Durant, R.~Mackenzie, and J.~F. Cooper, ``Towards automated
  analysis for neutron reflectivity,'' {\em Machine Learning: Science and
  Technology}, vol.~2, no.~3, p.~035006, 2021.

\bibitem{loaiza2021towards}
J.~M.~C. Loaiza and Z.~Raza, ``Towards reflectivity profile inversion through
  artificial neural networks,'' {\em Machine Learning: Science and Technology},
  vol.~2, no.~2, p.~025034, 2021.

\bibitem{doucet2021machine}
M.~Doucet, R.~K. Archibald, and W.~T. Heller, ``Machine learning for neutron
  reflectometry data analysis of two-layer thin films,'' {\em Machine Learning:
  Science and Technology}, vol.~2, no.~3, p.~035001, 2021.

\bibitem{aoki2021deep}
H.~Aoki, Y.~Liu, and T.~Yamashita, ``Deep learning approach for an interface
  structure analysis with a large statistical noise in neutron reflectometry,''
  {\em Scientific reports}, vol.~11, no.~1, pp.~1--9, 2021.

\bibitem{lee2016direct}
C.~Lee, F.~Katmis, P.~Jarillo-Herrero, J.~S. Moodera, and N.~Gedik, ``Direct
  measurement of proximity-induced magnetism at the interface between a
  topological insulator and a ferromagnet,'' {\em Nature Communications},
  vol.~7, no.~1, pp.~1--6, 2016.

\bibitem{kingma2013auto}
D.~P. Kingma and M.~Welling, ``Auto-encoding variational bayes,'' {\em arXiv
  preprint arXiv:1312.6114}, 2013.

\bibitem{higgins2016beta}
I.~Higgins, L.~Matthey, A.~Pal, C.~Burgess, X.~Glorot, M.~Botvinick,
  S.~Mohamed, and A.~Lerchner, ``$\beta$-{VAE}: Learning basic visual concepts
  with a constrained variational framework,'' {\em Proc. ICLR}, 2017.

\bibitem{ml-pnr}
N.~Andrejevic and Z.~Chen, ``Repository for machine learning-assisted analysis
  of polarized neutron reflectometry measurements.,'' 2021.

\bibitem{nogueira2012fluctuation}
F.~S. Nogueira and I.~Eremin, ``Fluctuation-induced magnetization dynamics and
  criticality at the interface of a topological insulator with a magnetically
  ordered layer,'' {\em Physical Review Letters}, vol.~109, no.~23, p.~237203,
  2012.

\bibitem{luo2013massive}
W.~Luo and X.-L. Qi, ``Massive dirac surface states in topological
  insulator/magnetic insulator heterostructures,'' {\em Physical Review B},
  vol.~87, no.~8, p.~085431, 2013.

\bibitem{wang2019observation}
F.~Wang, D.~Xiao, W.~Yuan, J.~Jiang, Y.-F. Zhao, L.~Zhang, Y.~Yao, W.~Liu,
  Z.~Zhang, C.~Liu, {\em et~al.}, ``Observation of interfacial
  antiferromagnetic coupling between magnetic topological insulator and
  antiferromagnetic insulator,'' {\em Nano Letters}, vol.~19, no.~5,
  pp.~2945--2952, 2019.

\bibitem{lauter2009highlights}
V.~Lauter, H.~Ambaye, R.~Goyette, W.-T.~H. Lee, and A.~Parizzi, ``Highlights
  from the magnetism reflectometer at the sns,'' {\em Physica B: Condensed
  Matter}, vol.~404, no.~17, pp.~2543--2546, 2009.

\bibitem{NEURIPS2019_9015}
A.~Paszke, S.~Gross, F.~Massa, A.~Lerer, J.~Bradbury, G.~Chanan, T.~Killeen,
  Z.~Lin, N.~Gimelshein, L.~Antiga, A.~Desmaison, A.~Kopf, E.~Yang, Z.~DeVito,
  M.~Raison, A.~Tejani, S.~Chilamkurthy, B.~Steiner, L.~Fang, J.~Bai, and
  S.~Chintala, ``Pytorch: An imperative style, high-performance deep learning
  library,'' in {\em Advances in Neural Information Processing Systems 32}
  (H.~Wallach, H.~Larochelle, A.~Beygelzimer, F.~d\textquotesingle
  Alch\'{e}-Buc, E.~Fox, and R.~Garnett, eds.), pp.~8024--8035, Curran
  Associates, Inc., 2019.

\bibitem{zhao2019variational}
Q.~Zhao, E.~Adeli, N.~Honnorat, T.~Leng, and K.~M. Pohl, ``Variational
  autoencoder for regression: Application to brain aging analysis,'' in {\em
  International Conference on Medical Image Computing and Computer-Assisted
  Intervention}, pp.~823--831, Springer, 2019.

\end{thebibliography}

% Supplementary Material
\clearpage
\setcounter{section}{0}
\setcounter{equation}{0}
\setcounter{figure}{0}
\setcounter{table}{0}
\setcounter{page}{1}
\makeatletter
\renewcommand{\theequation}{S\arabic{equation}}
\renewcommand{\thefigure}{S\arabic{figure}}

\onecolumngrid
\begin{center}
    \large\textbf{Supplementary material for ``Elucidating proximity magnetism through polarized neutron reflectometry and machine learning''} \\
    \vspace{\baselineskip}
    \normalsize Nina Andrejevic,\textit{$^{1,2,*,\dagger}$} Zhantao Chen,\textit{$^{1,3,*}$} Thanh Nguyen,\textit{$^{1,4}$}
    Leon Fan,\textit{$^5$} Henry Heiberger,\textit{$^{5}$} \\ Ling-Jie Zhou,\textit{$^{6}$} Yi-Fan Zhao,\textit{$^{6}$} Cui-Zu Chang,\textit{$^{6}$} Alexander Grutter,\textit{$^{7}$} and Mingda Li \textit{$^{1,4,\dagger}$} \\
    \textsuperscript{1}\textit{Quantum Measurement Group, Massachusetts Institute of Technology, Cambridge, MA 02139, USA} \\
    \textsuperscript{2}\textit{Department of Materials Science and Engineering,} \\
    \textit{Massachusetts Institute of Technology, Cambridge, MA 02139, USA} \\
    \textsuperscript{3}\textit{Department of Mechanical Engineering, Massachusetts Institute of Technology, Cambridge, MA 02139, USA} \\
    \textsuperscript{4}\textit{Department of Nuclear Science and Engineering,} \\
    \textit{Massachusetts Institute of Technology, Cambridge, MA 02139, USA} \\
    \textsuperscript{5}\textit{Department of Electrical Engineering and Computer Science,} \\
    \textit{Massachusetts Institute of Technology, Cambridge, MA 02139, USA} \\
    \textsuperscript{6}\textit{Department of Physics, The Pennsylvania State University, University Park, PA 16802, USA} \\
    \textsuperscript{7}\textit{National Institute of Standards and Technology,} \\
    \textit{Center for Neutron Research, Gaithersburg, MD 20899, USA} \\
    
\end{center}
\vfill

{\color{revision}\section*{Parameter retrieval with GenX}}

\begin{figure*}[t]
    \centering
    \includegraphics[width=0.9\linewidth]{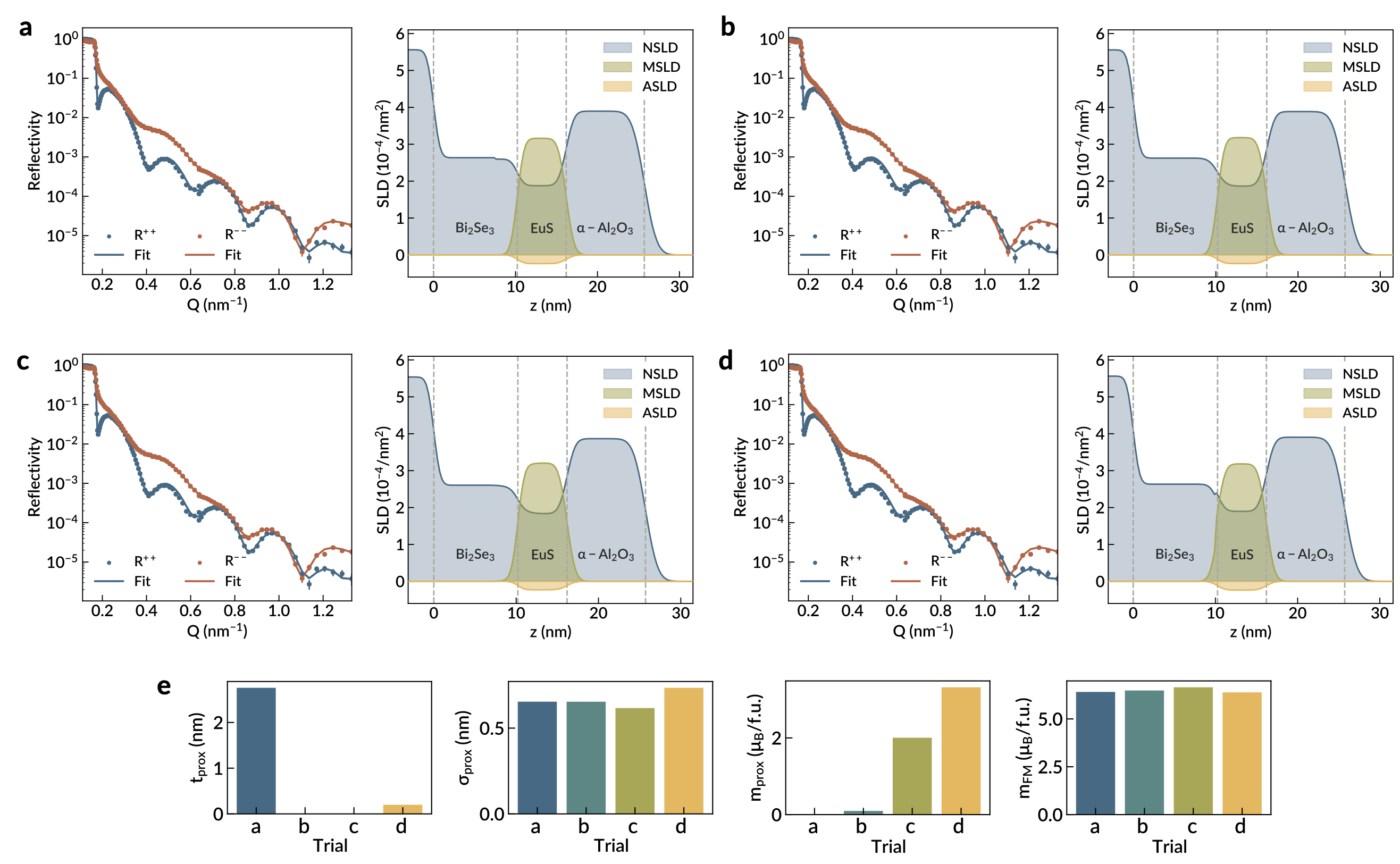}
    \caption{{\color{revision}\textbf{Representative GenX fits with varied initialization.} \textbf{a.} (Left) Reflectometry profile of the Bi$_2$Se$_3$/EuS system measured at 5 K. Solid lines correspond to the fit obtained using the GenX paramter refinement program with a particular initial parameter population. Error-bars represent ±1 standard deviation. (Right) Nuclear, magnetic, and absorption SLD profiles corresponding to the fit in the left panel. \textbf{b-c.} Analogous to \textbf{a} but obtained using different initial populations for the GenX fit. \textbf{e.} Fitted values for selected parameters obtained from the fits displayed in \textbf{a}-\textbf{d}.}}
    \label{fig:FigS1}
\end{figure*}

{\color{revision}We illustrate some of the challenges that may be encountered using conventional parameter refinement programs when the sample parameters cannot be adequately constrained using additional characterization methods. Figs. \ref{fig:FigS1}a-d show the fitted reflectivity and corresponding SLD profiles of the 5 K PNR measurement of Bi$_2$Se$_3$/EuS obtained through the GenX parameter refinement program using different initial populations. While all four fits are of comparable quality (Table \ref{tab:fom}) and generate similar SLD profiles, the underlying sample parameters differ considerably, as shown in Fig. \ref{fig:FigS1}e. In particular, while the interface roughness between the Bi$_2$Se$_3$ and EuS films and the FM film thickness are comparable for all four fits, the thickness and magnetization of the proximity layer at the interface vary dramatically; in fact, the fits shown in Figs. \ref{fig:FigS1}a-b correspond to little to no proximity magnetism, in contrast to the experimentally-determined result. This motivates the development of an alternate approach to resolve potential ambiguities resulting from insufficient parameter constraints or expert supervision during fitting.} \\

\begin{table}[h]
    \footnotesize
    \renewcommand{\arraystretch}{1.2}
	\centering
	\caption{GenX fit quality with different initial populations.}
	\begin{threeparttable}
	\begin{tabular}{@{}p{0.5\textwidth}@{}}
	\centering
	\begin{tabular}{ccc}
		\hline\hline
		Trial & Figure of merit (FoM)\tnote{*} & MSE \\
		\hline
		1 & $7.288 \times 10^{-4}$ & $1.032 \times 10^{-4}$ \\
        2 & $7.602 \times 10^{-4}$ & $1.034 \times 10^{-4}$ \\
        3 & $7.561 \times 10^{-4}$ & $1.144 \times 10^{-4}$ \\
        4 & $8.094 \times 10^{-4}$ & $1.006 \times 10^{-4}$ \\
		\bottomrule\addlinespace[1ex]
	\end{tabular}
	\end{tabular}
	\begin{tablenotes}\footnotesize
	    \item[*] The GenX FoM is the MSE between the logarithms of the true and fitted reflectivity profiles, added over the spin channels.
	\end{tablenotes}
	\end{threeparttable}
	\label{tab:fom}
\end{table}

\section*{Motivating example for VAE-based parameter retrieval}
\begin{figure*}[t]
    \centering
    \includegraphics[width=0.9\linewidth]{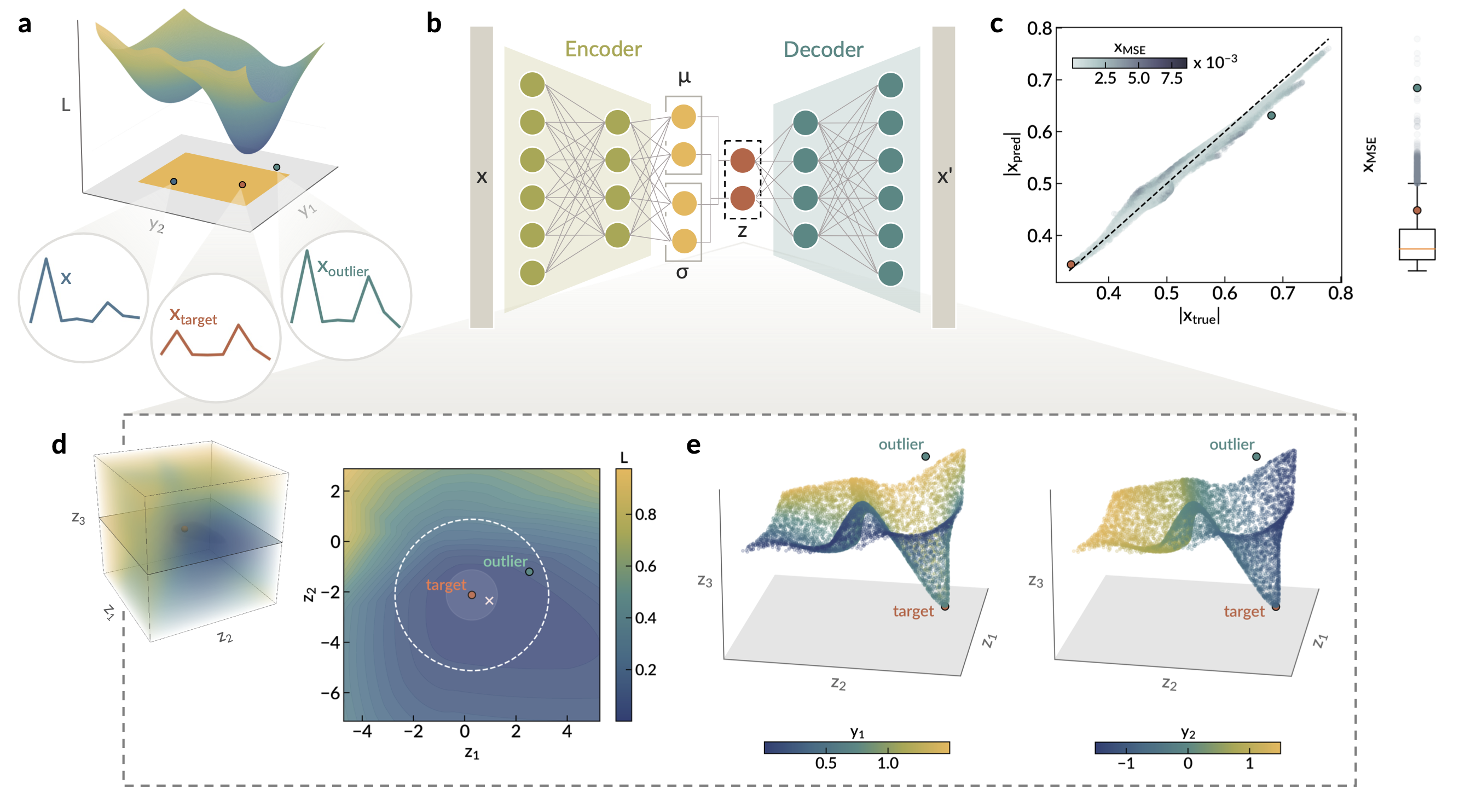}
    \caption{\textbf{Advantages of VAE-based parameter retrieval.} \textbf{a.} Toy model of a signature $\textbf{x}$ generated by two parameters, $y_1$ and $y_2$. The loss $L$ denoting the MSE between an arbitrary $\textbf{x}$ and a target signature $\textbf{x}_\text{target}$ contains one global and one local minimum in $(y_1, y_2)$-space. \textbf{b.} Schematic illustration of the VAE architecture. \textbf{c.} Norms of the predicted versus true signatures of the test dataset. Each point is colored according to the MSE between the associated $\textbf{x}_\text{pred}$ and $\textbf{x}_\text{true}$. The signature $\textbf{x}_\text{target}$ (red point) generated by $(y_1, y_2)$ within the domain of training samples (yellow patch in \textbf{a}) is reconstructed with comparatively low MSE, while the signature $\textbf{x}_\text{outlier}$ outside the domain (teal point) is an outlier, reflected also in the adjacent box plot. \textbf{d.} Density plot of the loss $L$ as a function of the three-dimensional latent space $\textbf{z}$ showing a single minimum in the region of interest. A two dimensional slice through the $z_3=0$ plane is shown at the right. The latent representations of the target and outlier signatures from \textbf{a} projected onto the $z_3=0$ plane are indicated by the red and teal points, respectively. The shaded disk and dashed circle denote one and three standard deviations away from the latent coordinate of the target signature, respectively. The ``$\times$" denotes the minimum of the loss $L$. \textbf{e.} Visualization of the latent representation of the test dataset. Each point is colored according to the associated true values of $y_1$ (left) and $y_2$ (right), showing that $z_1$ and $z_2$ vary jointly with $y_1$ and $y_2$, respectively. The latent representations of the target and outlier signatures are indicated by the red and teal points, respectively.}
    \label{fig:FigS2}
\end{figure*}

To demonstrate the potential advantages of VAE-based parameter retrieval, we consider the simplified example illustrated in Fig. \ref{fig:FigS2}. The objective is to determine the sample parameters $y_1$ and $y_2$ that best reproduce the comparatively high-dimensional target signature $\textbf{x}_\text{target}$, denoted in red in Fig. \ref{fig:FigS2}a. In this simple example, the landscape of a mean squared error (MSE) loss $L = \|\textbf{x} - \textbf{x}_\text{target}\|^2$ contains one global and one local minimum in ($y_1, y_2$)-space; thus, rather than optimize directly in this space, we train the VAE to recover the input from a low (3)-dimensional latent representation, $\textbf{z}$ (Fig. \ref{fig:FigS2}b). Since the latent space is organized such that nearby points correspond to similar reconstructions of the input, the loss function viewed in this space has only a single, global minimum in the region of interest (Fig. \ref{fig:FigS2}d). {\color{revision}We note that the dimension of $\textbf{z}$ must be higher than that of ($y_1, y_2$), \textit{i.e.}, greater than 2, but is termed low-dimensional as it is substantially smaller than the dimension of $\textbf{x}$. This is particularly true for the target application of PNR, where measured reflectometry profiles contain more than 100 points which must be correlated to only a dozen or so underlying sample parameters.} Additionally, if the target signature were instead an outlier of the chosen domain for parameters $y_1$ and $y_2$ (teal point in Fig. \ref{fig:FigS2}a), we could readily single it out by its large reconstruction error, shown in Fig. \ref{fig:FigS2}c. {\color{revision}This would signify a need to expand or modify the parameter space, thus informing the suitability of the chosen parameter ranges. For example, the latent vector corresponding to the outlier in Fig. \ref{fig:FigS2}c is seen to sit off the manifold generated by mapping the test set to the latent space, shown in Fig. \ref{fig:FigS2}e, indicating it is not generated by parameters in the chosen range.} Thus far, the considered framework is fully unsupervised, with no direct correspondence between the intermediate features and the original parameters $y_1$ and $y_2$. However, as we expect the output to evolve predictably with the sample parameters, we can supervise the VAE training by conditioning a subset of the latent features $\textbf{z}$ to vary directly with $y_1$ and $y_2$ following the formulation of Ref. \citenum{zhao2019variational}, as shown in Fig. \ref{fig:FigS2}e. This can make the latent space more informative and interpretable by encouraging each selected dimension to organize according to a specific physical parameter, thereby facilitating quantitative prediction of its value.

\begin{figure*}[!t]
    \centering
    \includegraphics[width=0.8\linewidth]{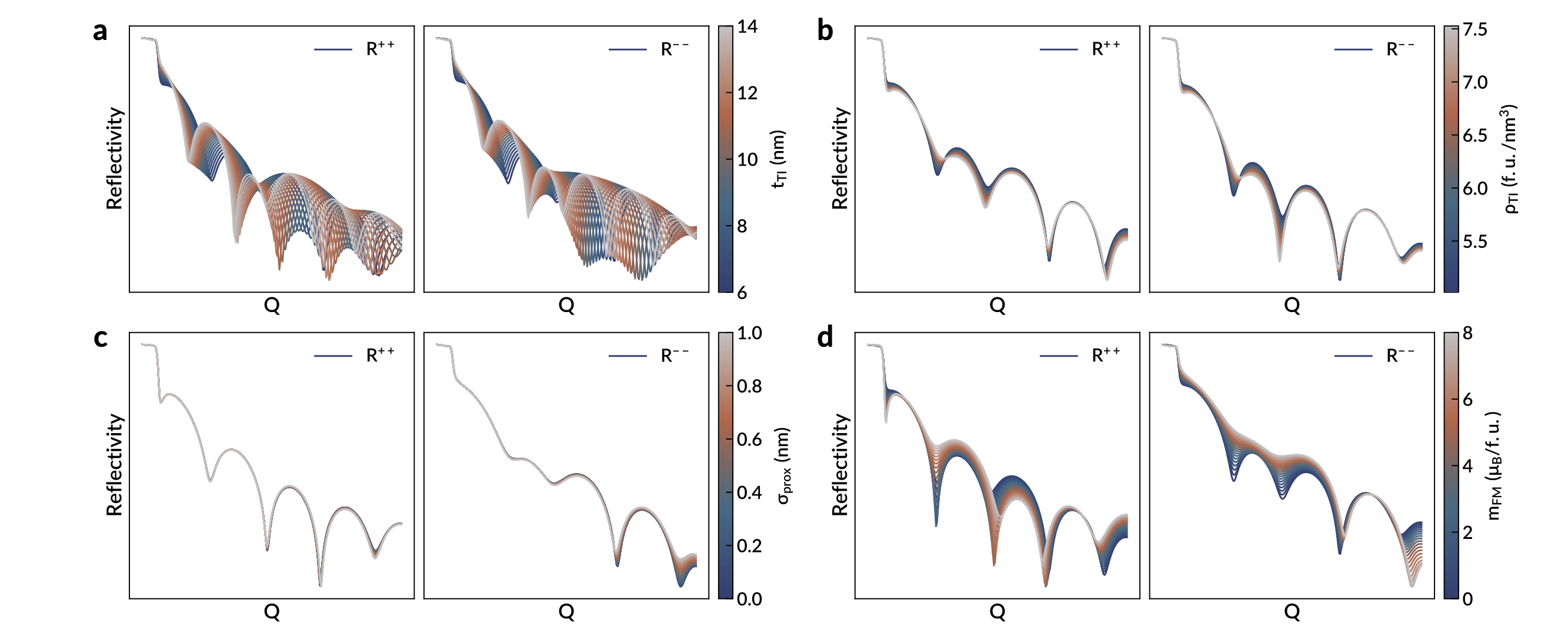}
    \caption{\textbf{PNR profile evolution with sample parameters.} Representative examples for the evolution of PNR profiles with respect to a continuous change in \textbf{a.} TI layer thickness, \textbf{b.} TI layer density, \textbf{c.} top surface roughness of the interfacial proximity layer, and \textbf{d.} magnetization strength of the FM layer.}
    \label{fig:FigS3}
\end{figure*}

\begin{figure*}[!h]
    \centering
    \includegraphics[width=0.9\linewidth]{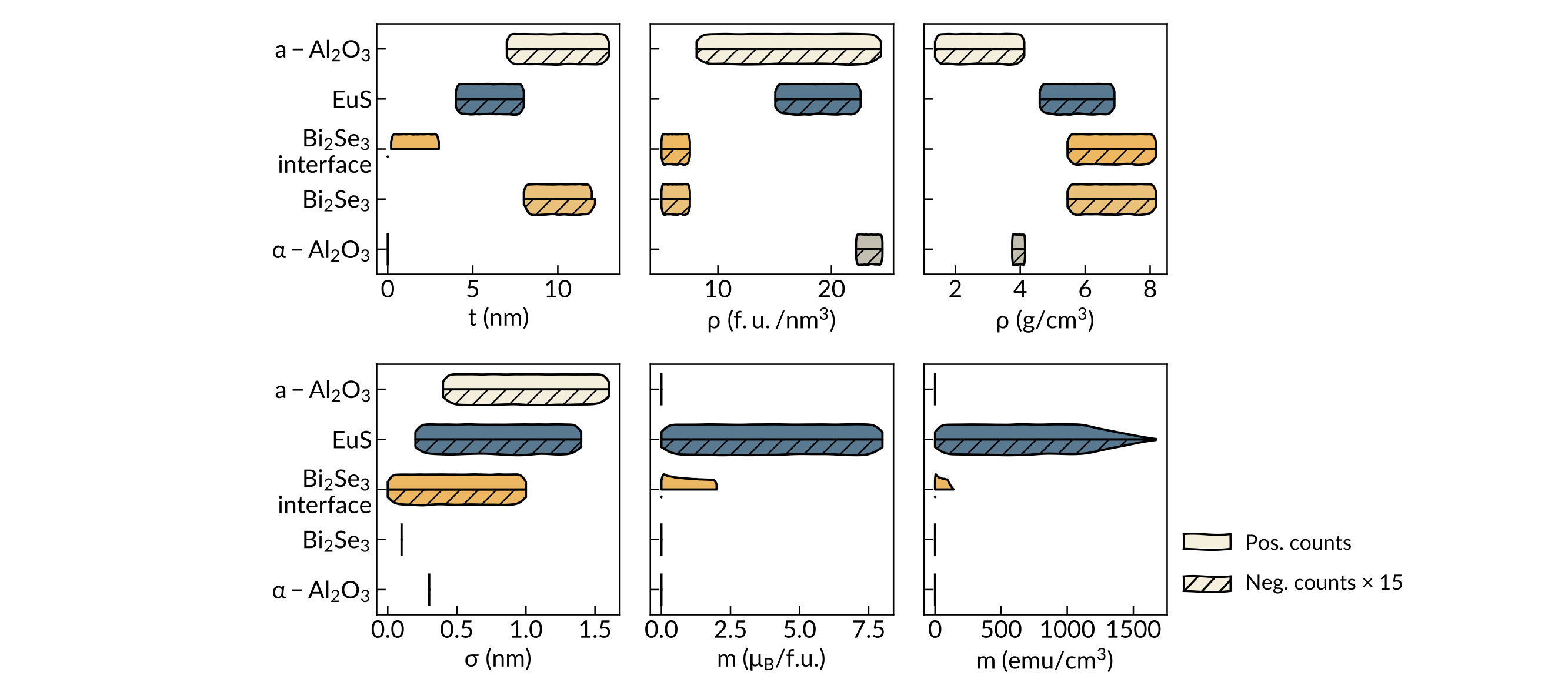}
    \caption{\textbf{Parameter ranges for Bi$_2$Se$_3$/EuS data generation.} Distribution of parameter values for the density, thickness, interface roughness, and magnetization of each layer in the generated data. Distributions are split between samples with (solid) and without (hatched) an interfacial proximity layer, {\color{revision}termed positive and negative counts in the legend, respectively}. Density and magnetization are sampled using their values in terms of formula units (middle column), which are compatible with the GenX simulation software. The corresponding values in conventional units are shown in the rightmost column (1 emu = 1 A$\cdot$m$^2$).}
    \label{fig:FigS4}
\end{figure*}

\begin{figure*}[!h]
    \centering
    \includegraphics[width=0.9\linewidth]{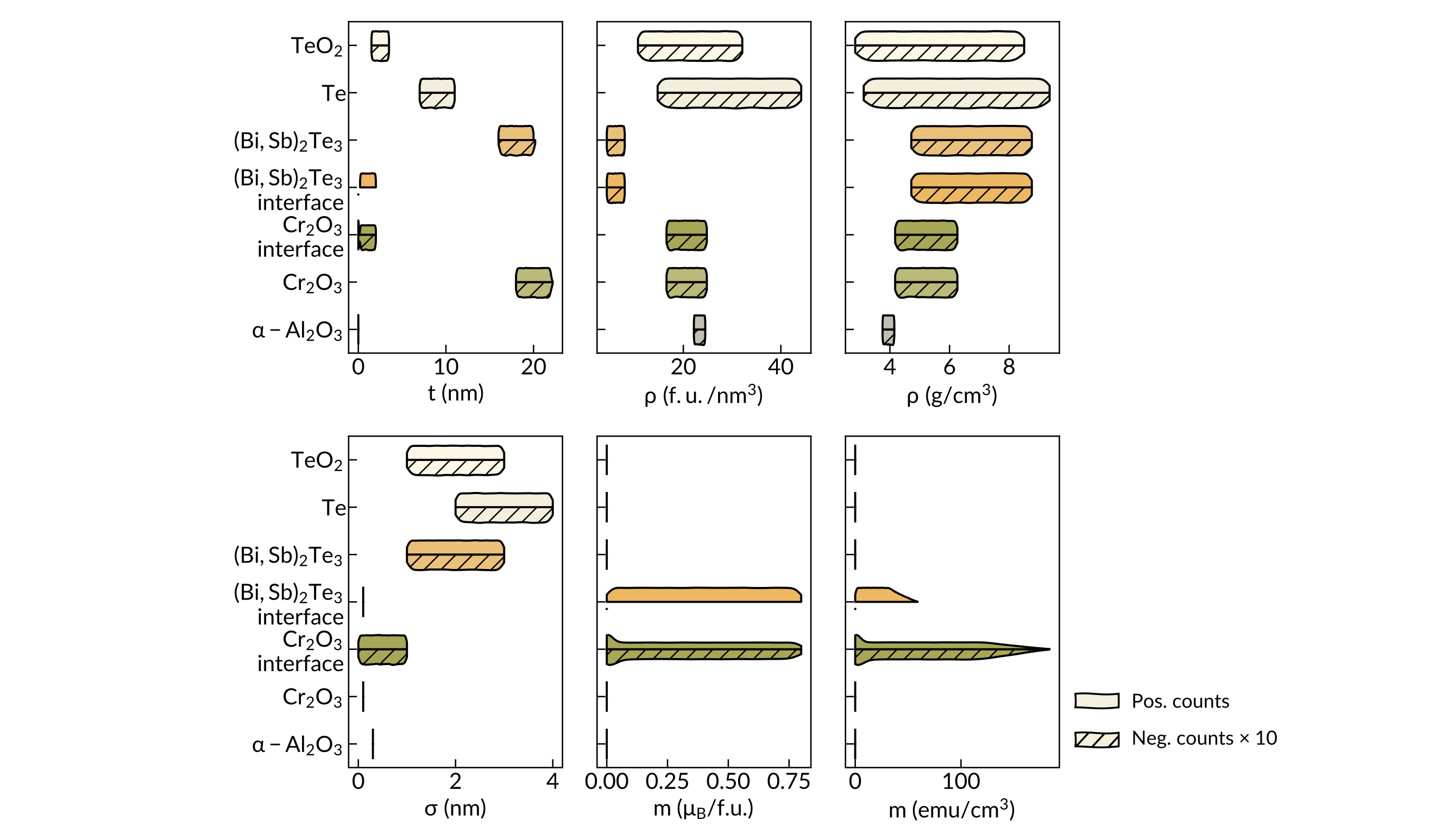}
    \caption{\textbf{Parameter ranges for (Bi,Sb)$_2$Te$_3$/Cr$_2$O$_3$ data generation.} Distribution of parameter values for the density, thickness, interface roughness, and magnetization of each layer in the generated data. Distributions are split between samples with (solid) and without (hatched) an interfacial proximity layer, {\color{revision}termed positive and negative counts in the legend, respectively}. Density and magnetization are sampled using their values in terms of formula units (middle column), which are compatible with the GenX simulation software. The corresponding values in conventional units are shown in the rightmost column (1 emu = 1 A$\cdot$m$^2$).}
    \label{fig:FigS5}
\end{figure*}

\begin{figure*}[!h]
    \centering
    \includegraphics[width=0.9\linewidth]{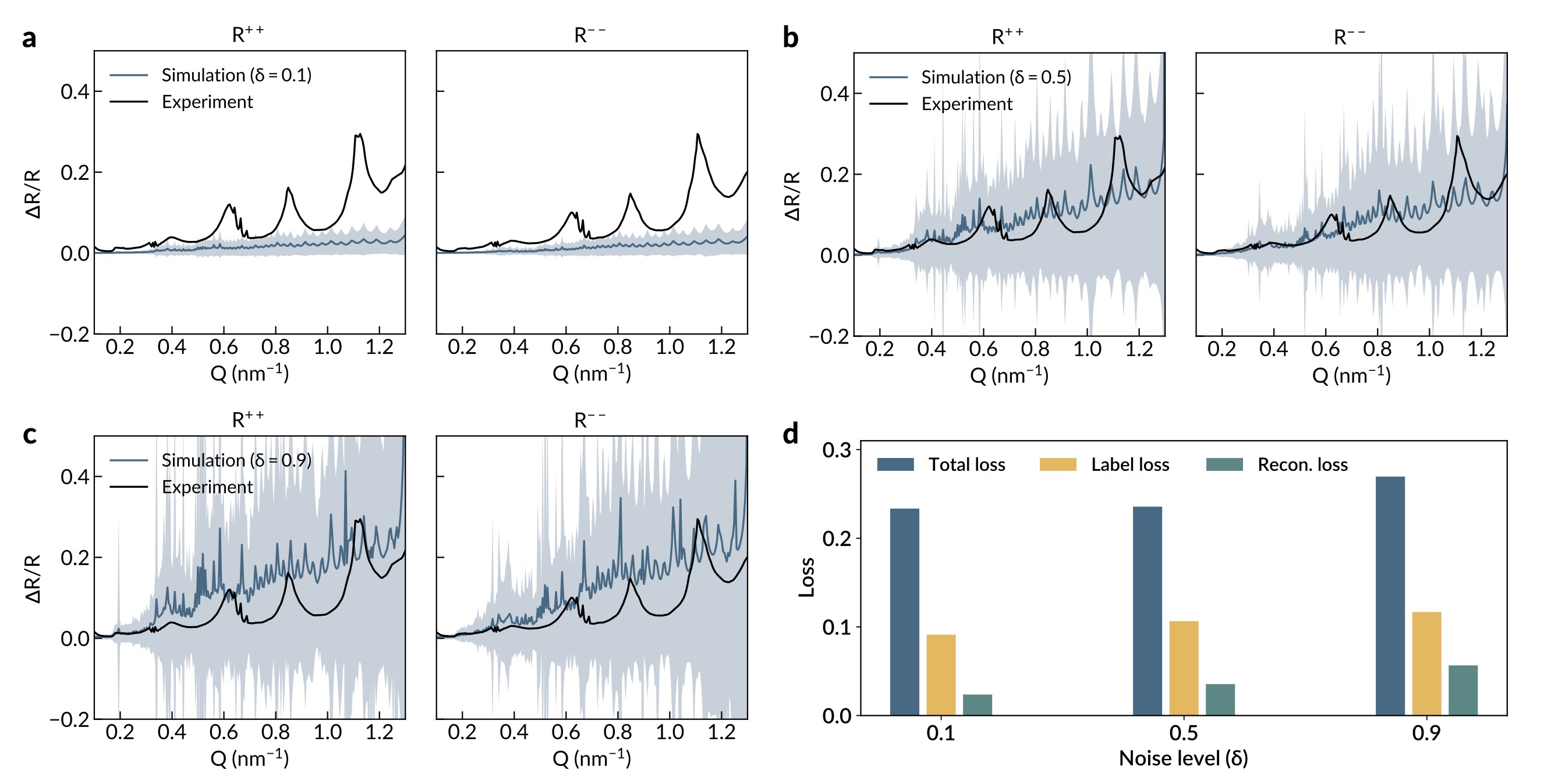}
    \caption{{\color{revision}\textbf{Selection of the synthetic noise level $\boldsymbol \delta$.} \textbf{a-c.} Comparison between simulated and experimental $\Delta R(Q)/R(Q)$ for \textbf{a.} $\delta = 0.1$, \textbf{b.} $\delta = 0.5$, and \textbf{c.} $\delta = 0.9$ for each spin channel. Solid black line indicates the experimental value while the solid blue line shows the average of 1000 simulated values. Light blue shading indicates one standard deviation from the average simulated value. \textbf{d.} Total, reconstruction, and label losses achieved by identical VAEs trained on synthetic PNR profiles perturbed by the different noise levels.}}
    \label{fig:FigS6}
\end{figure*}

\begin{figure*}[!h]
    \centering
    \includegraphics[width=0.9\linewidth]{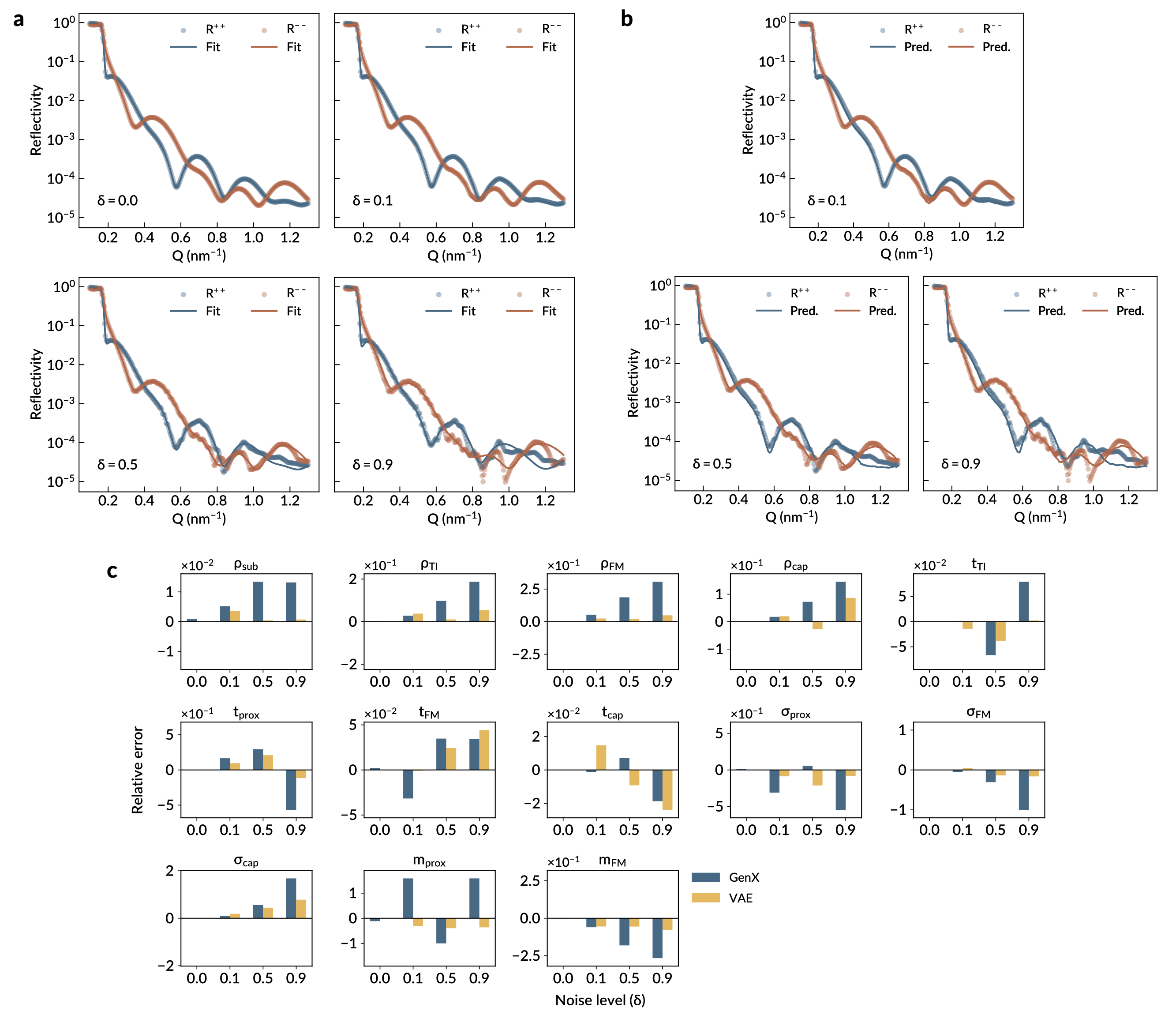}
    \caption{{\color{revision}\textbf{Noise dependence of conventional and VAE-based methods.} \textbf{a.} Simulated (points) and fitted (solid lines) reflectivity profiles with different levels of noise, indicated by the parameter $\delta$, using GenX parameter refinement. The simulated curve is one of the Bi$_2$Se$_3$/EuS training examples. \textbf{b.} Simulated (points) and reconstructed (solid lines) reflectivity profiles with different levels of noise, indicated by the parameter $\delta$, using the VAE framework. \textbf{c.} Fitted or predicted parameter values corresponding to the profiles in \textbf{a} and \textbf{b} obtained using GenX (blue) and the VAE framework (yellow).}}
    \label{fig:FigS7}
\end{figure*}

{\color{revision}\section*{Data preparation and noise estimation}}
{\color{revision}As described in the main text, training data are generated using the GenX neutron reflectivity modeling code \cite{bjorck2007genx}, which simulates the PNR profiles using a layered representation of the candidate heterostructures parameterized by the density, thickness, roughness, and magnetization of each layer. The variable parameters are sampled uniformly at random over a range of experimentally feasible values, shown in Figs. \ref{fig:FigS4} and \ref{fig:FigS5} for the Bi$_2$Se$_3$/EuS and (Bi,Sb)$_2$Te$_3$/Cr$_2$O$_3$ systems, respectively. Fig. \ref{fig:FigS3} shows the representative evolution of the PNR profiles with respect to selected parameters for the Bi$_2$Se$_3$/EuS system.
For noise estimation, it is reasonable to assume that the uncertainties in the measured reflectometry profiles follow the Poisson distribution, \textit{i.e.} the uncertainty in $N(Q)$ detected counts is approximately $\Delta N(Q) = \sqrt{N(Q)}$. $N(Q)$ relates to the normalized counting statistics, $R(Q)$, according to $R(Q) = N(Q)/N_0$, where $N_0$ is the maximum count rate, and hence,}
\begin{equation*}
    {\color{revision}\Delta R(Q) = \frac{\Delta N(Q)}{N_0} = \frac{\sqrt{N(Q)}}{N_0} = \sqrt{\frac{R(Q)}{N_0}}.}
\end{equation*}
{\color{revision}As noted in the main text, each profile is simulated with a randomly sampled instrument background, \textit{i.e.} $R(Q) = R_n(Q) + R_\mathrm{bkg}$, where we define the neutron counts $N_n(Q) = N_0 R_n(Q)$ and background $N_\mathrm{bkg} = N_0 R_\mathrm{bkg}$. Then,}
\begin{equation*}
    {\color{revision}\frac{\Delta R(Q)}{R(Q)} = \delta \sqrt{\frac{R_\mathrm{bkg}}{R(Q)}},}
\end{equation*}
{\color{revision}where $\delta \equiv 1/\sqrt{N_{bkg}}$ is empirically determined such that $\Delta R(Q)/R(Q)$ approximates the ratio between experimental errorbars and data points. The simulated data are perturbed at each $Q$ by randomly sampling a Gaussian distribution with standard deviation $\sigma(Q) = \Delta R(Q)/R(Q)$, and computing the perturbed profile as $\tilde{R}(Q) = R(Q)(1 + \epsilon(Q))$, where $\epsilon(Q)$ denotes the sampled value. Independent perturbation of the profile at each $Q$ leads to a highly oscillatory profile, particularly at high $Q$ that does not emulate experiment well. Thus, the perturbed profiles are lightly smoothed by a uniform moving average over $\tilde{R}(Q)$ as a function of $1/Q$, which we find creates synthetic profiles that resemble experimental ones. Due to dimensionality reduction of the input profile by the VAE encoder, the VAE returns predicted profiles that are typically less noisy than the inputs (\textit{e.g.} see Fig. \ref{fig:FigS10}). Fig. \ref{fig:FigS6}a-c compares the simulated $\Delta R(Q)/R(Q)$ for different choices of $\delta$, 0.1, 0.5, and 0.9, against the experimental $\Delta R(Q)/R(Q)$ of the Bi$_2$Se$_3$/EuS sample. These results establish $\delta = 0.5$ as a suitable choice to best match the given experimental statistics. In Fig. \ref{fig:FigS6}d, we also plot the total loss, reconstruction loss, and label loss achieved by identical VAEs trained on synthetic PNR profiles perturbed with each noise level. Finally, we also evaluate the noise dependence of the conventional and VAE-based approaches. Fig. \ref{fig:FigS7}a shows fitted reflectivity profiles obtained using the GenX parameter refinement program for a simulated profile perturbed by different levels of noise. The case with $\delta = 0$ should in theory be perfectly recoverable as it is both simulated and fitted by the same program. We show the same profiles in Fig. \ref{fig:FigS7}b alongside reconstructions obtained by the VAE framework for $\delta > 0$. Table \ref{tab:noise} lists the MSEs between the fitted or predicted profile and the corresponding input obtained with each method for each noise level. While the MSEs of the VAE-predicted profiles are generally higher than those fitted by GenX (Table \ref{tab:noise}), the discrepancy is reduced at higher $\delta$, and at $\delta = 0.9$, the profiles predicted by the VAE are in fact a better match to the noise-free profiles than those obtained by GenX. Moreover, the parameter values predicted by the VAE almost always correspond much more closely with the true parameter values than those obtained from the GenX fit, as shown in Fig. \ref{fig:FigS7}c, which is particularly evident for larger values of $\delta$. Thus, the VAE may be particularly useful when working with data acquired in a short time, which often includes large statistical noise.}

\begin{table}[!b]
    \footnotesize
    \renewcommand{\arraystretch}{1.2}
	\centering
	\caption{Noise-dependence of MSE.}
	\begin{threeparttable}
	\begin{tabular*}{0.6\textwidth}{@{\extracolsep{\fill}} ccccc}
		\hline\hline
		& \multicolumn{2}{c}{GenX} & \multicolumn{2}{c}{VAE} \\
		\hline
		$\delta$ & MSE & MSE$_{\delta=0}$\tnote{*} & MSE & MSE$_{\delta=0}$ \\
		\hline
		0. & $6.81 \times 10^{-7}$ & $6.81 \times 10^{-7}$ & - & - \\
        0.1	& $3.72 \times 10^{-6}$ & $1.49 \times 10^{-6}$ & $2.05 \times 10^{-5}$ & $1.71 \times 10^{-5}$ \\
        0.5 & $4.21 \times 10^{-5}$ & $1.69 \times 10^{-5}$ & $7.63 \times 10^{-5}$ & $2.17 \times 10^{-5}$ \\
        0.9	& $1.39 \times 10^{-4}$ & $7.47 \times 10^{-5}$ & $1.72 \times 10^{-4}$ & $2.95 \times 10^{-5}$ \\
		\bottomrule\addlinespace[1ex]
	\end{tabular*}
	\begin{tablenotes}\footnotesize
	    \item[*] MSE$_{\delta=0}$ refers to the MSE between the fitted or predicted profile and the unperturbed simulation, the equivalent of setting $\delta=0$.
	\end{tablenotes}
	\end{threeparttable}
	\label{tab:noise}
\end{table}

\clearpage

\clearpage

\begin{figure*}[!h]
    \centering
    \includegraphics[width=0.95\linewidth]{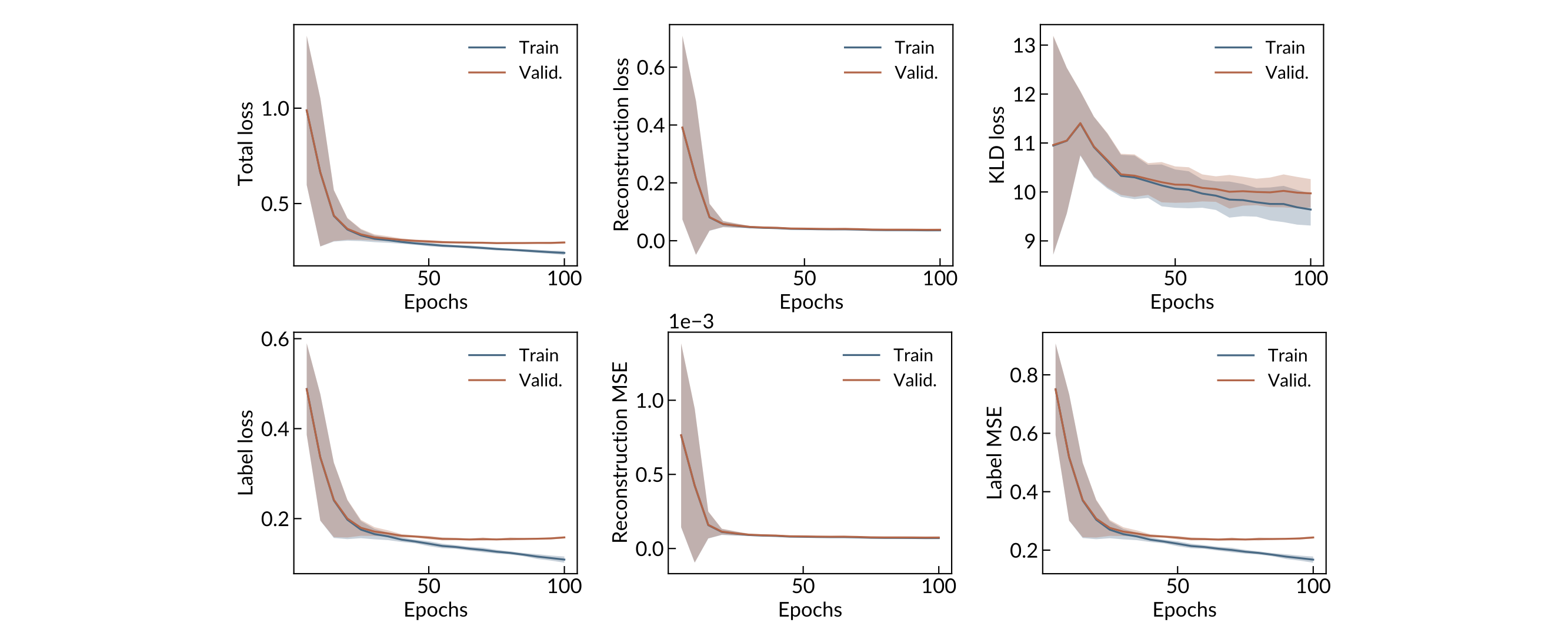}
    \caption{\textbf{Training history for the Bi$_2$Se$_3$/EuS system.} Trajectories of the total loss and individual contributions from reconstruction loss, KL divergence (KLD) loss, and label loss for the training and validation sets. The mean squared errors (MSEs) between the decoded and true PNR profiles (Reconstruction MSE) and between predicted and true sample parameters (Label MSE) are also shown.}
    \label{fig:FigS8}
\end{figure*}

\begin{figure*}[!hb]
    \centering
    \includegraphics[width=0.95\linewidth]{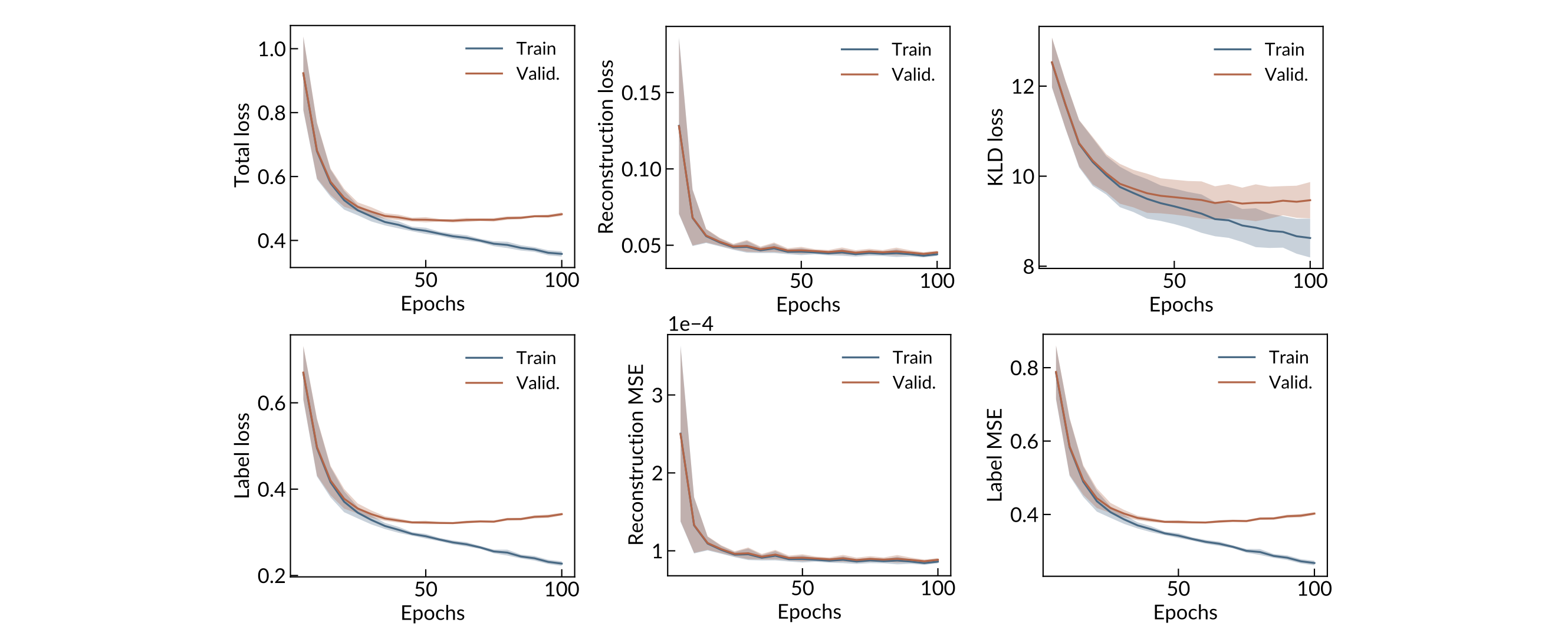}
    \caption{\textbf{Training history for the (Bi,Sb)$_2$Te$_3$/Cr$_2$O$_3$ system.} Trajectories of the total loss and individual contributions from reconstruction loss, KL divergence (KLD) loss, and label loss for the training and validation sets. The mean squared errors (MSEs) between the decoded and true PNR profiles (Reconstruction MSE) and between predicted and true sample parameters (Label MSE) are also shown.}
    \label{fig:FigS9}
\end{figure*}

\begin{figure*}[!h]
    \centering
    \includegraphics[width=0.9\linewidth]{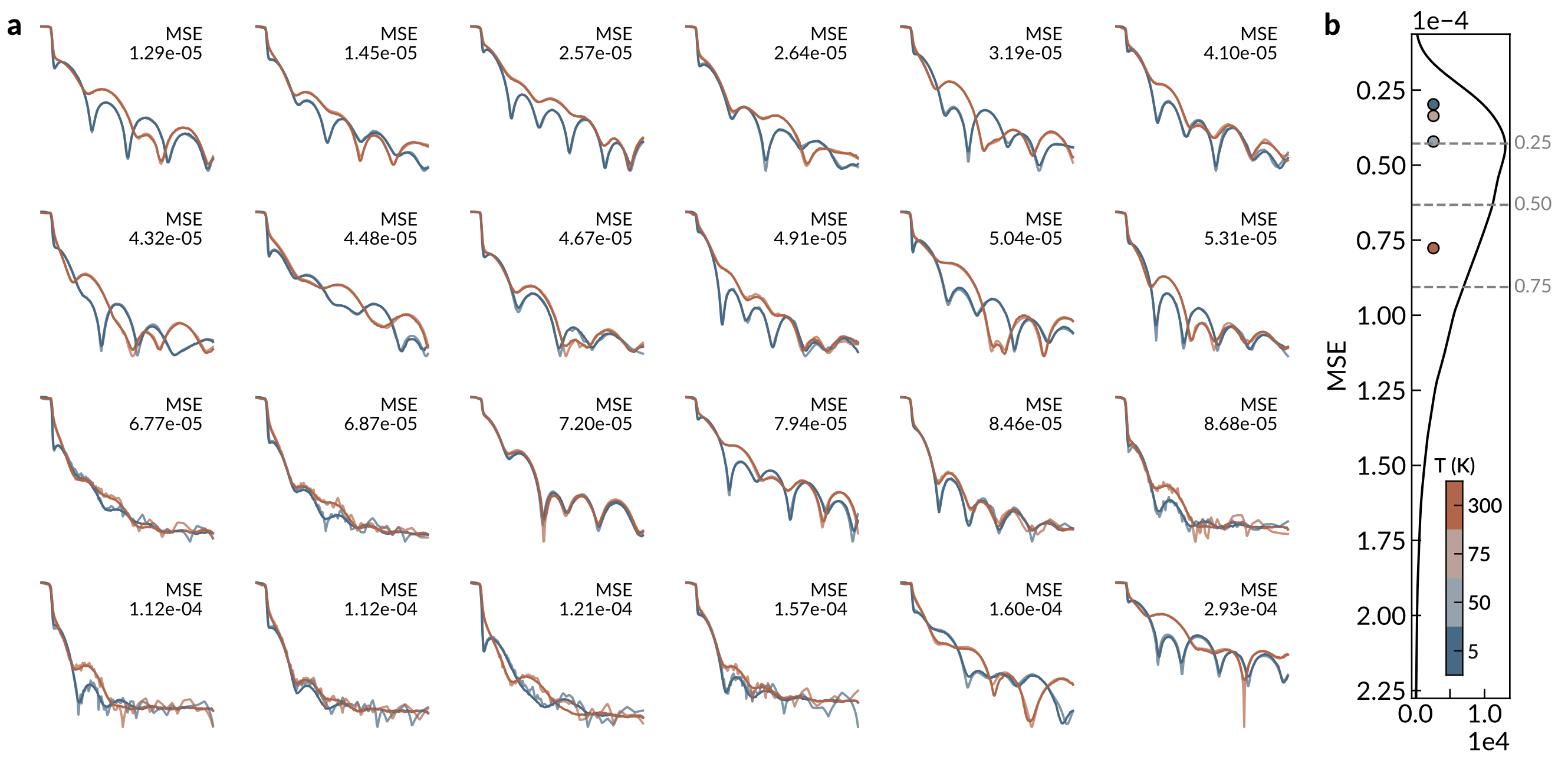}
    \caption{\textbf{VAE reconstruction accuracy for the Bi$_2$Se$_3$/EuS system.} \textbf{a.} Representative reconstructions of the test dataset within each error quartile, sorted from best (top row) to worst (bottom row). Predicted (true) reflectivities are plotted as {\color{revision} bold (light)} curves for each channel: $R^{++}$ (blue) and $R^{--}$ (red). The corresponding reconstruction error (MSE) is indicated in the top right corner of each subplot. \textbf{b.} Distribution of MSE values for the test dataset. Note the inverted y-axis of the plot, with the best performing quartile located at the top of the distribution. Quartiles are indicated and separated by dashed gray lines. Plotted points represent the reconstruction errors of the four experimental PNR profiles, colored by temperature.}
    \label{fig:FigS10}
\end{figure*}

\begin{figure*}[!hb]
    \centering
    \includegraphics[width=0.9\linewidth]{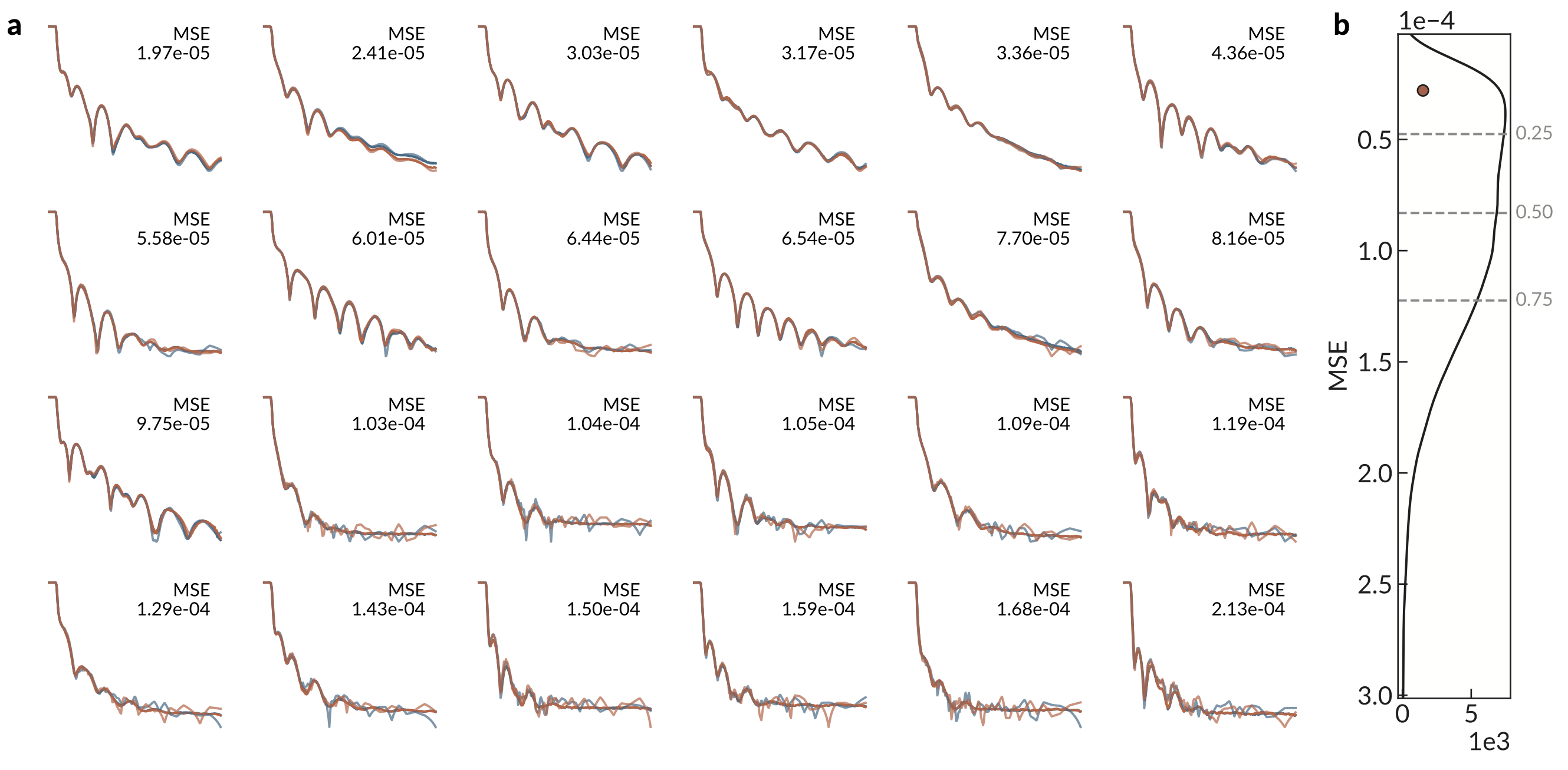}
    \caption{\textbf{VAE reconstruction accuracy for the (Bi,Sb)$_2$Te$_3$/Cr$_2$O$_3$ system.} \textbf{a.} Representative reconstructions of the test dataset within each error quartile, sorted from best (top row) to worst (bottom row). Predicted (true) reflectivities are plotted as {\color{revision} bold (light)} curves for each channel: $R^{++}$ (blue) and $R^{--}$ (red). The corresponding reconstruction error (MSE) is indicated in the top right corner of each subplot. \textbf{b.} Distribution of MSE values for the test dataset. Note the inverted y-axis of the plot, with the best performing quartile located at the top of the distribution. Quartiles are indicated and separated by dashed gray lines. The scattered point represents the reconstruction error of the experimental PNR profile, measured at 5 K.}
    \label{fig:FigS11}
\end{figure*}

\begin{figure*}[!h]
    \centering
    \includegraphics[width=0.8\linewidth]{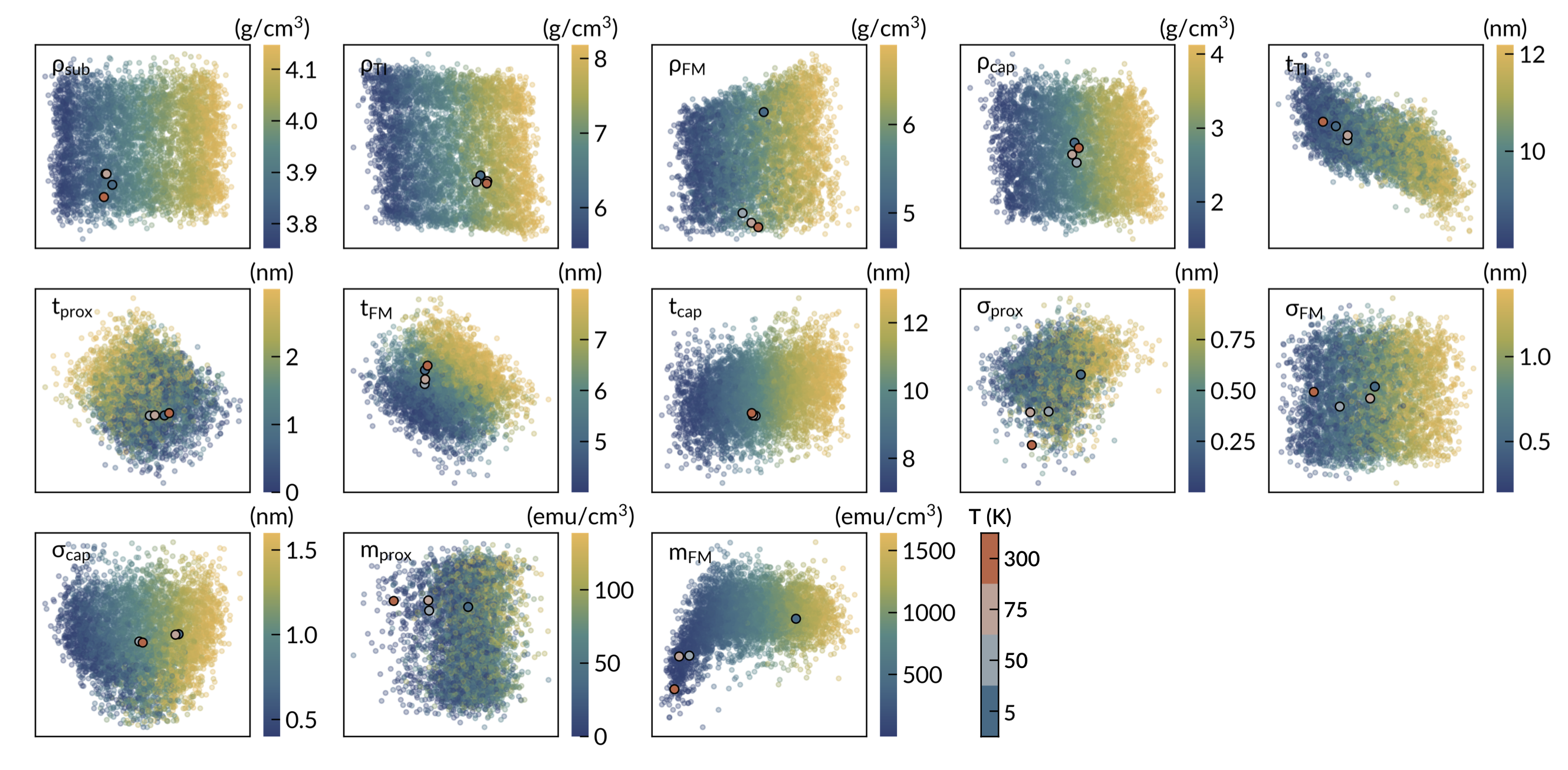}
    \caption{\textbf{Latent space visualization for the Bi$_2$Se$_3$/EuS system.} Projections of the latent encoding of the test dataset along the directions of largest gradient for different sample parameters (density $\rho$, thickness $t$, {\color{revision}roughness $\sigma$}, and magnetization $m$), colored by their true values. Outlined points show the latent encoding of the four experimental measurements.}
    \label{fig:FigS12}
\end{figure*}

\begin{figure*}[!hb]
    \centering
    \includegraphics[width=0.8\linewidth]{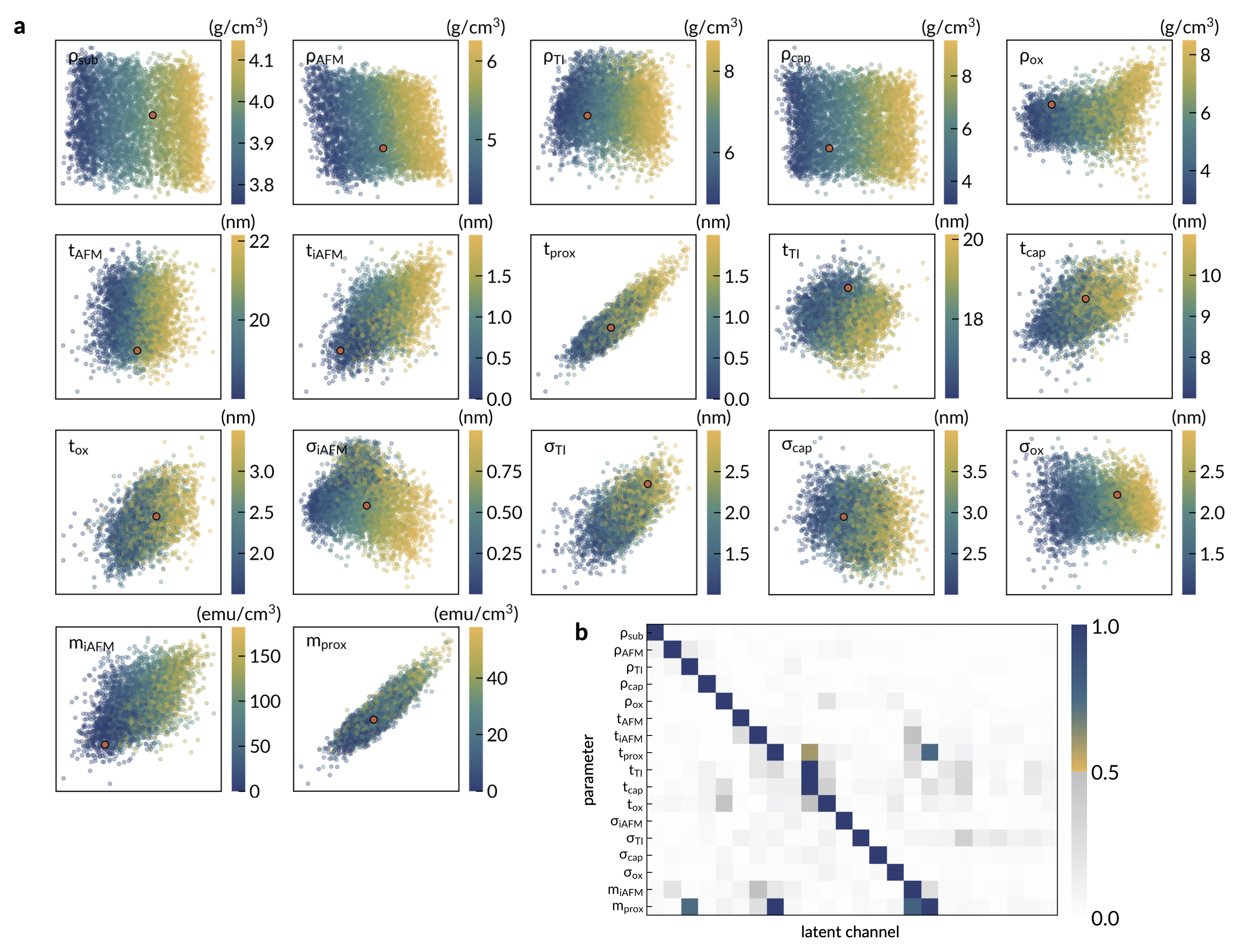}
    \caption{\textbf{Latent space visualization for the (Bi,Sb)$_2$Te$_3$/Cr$_2$O$_3$ system.} \textbf{a.} Projections of the latent encoding of the test dataset along the directions of largest gradient for different sample parameters (density $\rho$, thickness $t$, {\color{revision}roughness $\sigma$}, magnetization $m$) and proximity classification, colored by their true values. The red, outlined point shows the latent encoding of the experimental measurement. \textbf{b.} Parameter entanglement inferred from gradients along each latent channel. Heatmap indicates the relative magnitudes of the gradients of a given parameter with respect to each latent channel.}
    \label{fig:FigS13}
\end{figure*}

{\color{revision}\section*{VAE training details}}
We train the VAE by minimizing the loss,
\begin{equation*}
    \mathcal{L} = - \mathbb{E}_{\textbf{z} \sim \textbf{q}_\theta(\mathbf{z} \mid \mathbf{x})}[\log p_\phi(\mathbf{x}\!\mid\!\mathbf{z})] + \beta D_{KL}(q_\theta(\mathbf{z}\mid\!\mathbf{x})\!\mid\mid\!p(\textbf{z})) + \lambda \|\textbf{v} - \textbf{v}'\|^2.
\end{equation*}
{\color{revision}The first term represents the reconstruction loss, or the expected negative log-likelihood of $\textbf{x}$ given $\textbf{z}$; the second term regularizes the distribution of latent features; and the third gives the error between the (standardized) true and predicted parameter values of the regressor. The latent channels which are conditioned on specific sample parameters are drawn from Gaussian prior distributions with mean $\textbf{v}$ and variance 1, \textit{i.e.}, $p(z_i) \sim \mathcal{N}(v_i,1)$ for each parameter $i$, while the so-called free latent channels adopt the conventional Gaussian prior with zero mean, \textit{i.e.}, $p(z_i) \sim \mathcal{N}(0,1)$. The approximate posterior $q_\theta(z\mid\!x)$ and likelihood $p_\phi(x\!\mid\!z)$ are taken to be Gaussian-distributed; for this choice of likelihood, minimizing the negative log-likelihood of $\textbf{x}$ with respect to $\theta$ is equivalent to minimizing the mean squared error between $x$ and $\textbf{x}'$, where $\textbf{x}'$ represents the reconstructed input $\textbf{x}$. Thus, the first term is directly implemented as $\|\textbf{x} - \textbf{x}'\|^2$. The encoder, decoder, and regressor networks were optimized jointly using the Adam optimizer; the hyperparameters used for the final model are listed in Table \ref{tab:hyperparameters}. The training histories for the Bi$_2$Se$_3$/EuS and (Bi,Sb)$_2$Te$_3$/Cr$_2$O$_3$ are shown in Figs. \ref{fig:FigS8} and \ref{fig:FigS9}, respectively.}

\begin{table}[b]
    \footnotesize
    \renewcommand{\arraystretch}{1.2}
	\centering
	\caption{Final network hyperparameters.}
	\begin{threeparttable}
	\begin{tabular}{cc}
		\hline\hline
		Hyperparameter & Value \\
		\hline
		Learning rate & $1 \times 10^{-3}$ \\
		Batch size & 512 \\
		Number of convolutional layers & 2 \\
		Number of dense layers & 4 \\
		Latent dimension & 24 \\
		Number of starting filters & 16 \\
		Convolution kernel size & 7 \\
		Maxpool size & 4 \\
		$\beta$ & 0.01 \\
		$\lambda$ & 0.05 \\
		\bottomrule\addlinespace[1ex]
	\end{tabular}
	\end{threeparttable}
	\label{tab:hyperparameters}
\end{table}

\clearpage
{\color{revision}\section*{VAE Reconstruction accuracy}}

\vspace{-0.5cm}

\clearpage
{\color{revision}\section*{Latent space visualization}}

\vspace{-0.5cm}

\clearpage
{\color{revision}\section*{VAE regressor performance}}
\begin{figure*}[b]
    \centering
    \includegraphics[width=0.9\linewidth]{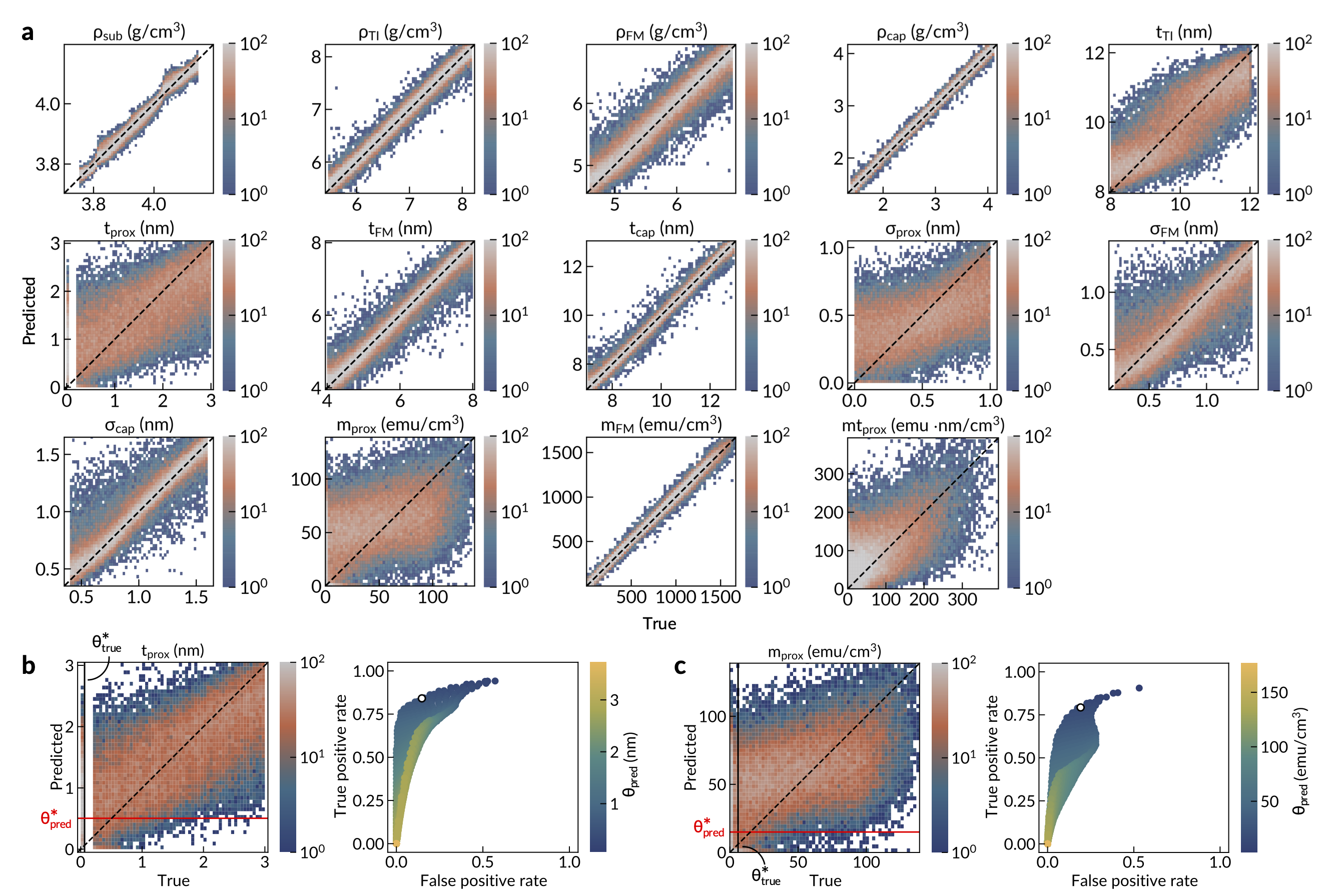}
    \caption{\textbf{Regressor performance for the Bi$_2$Se$_3$/EuS system.} \textbf{a.} Histograms of predicted versus true values for different sample parameters of the test dataset. {\color{revision}\textbf{b-c.} Threshold determination for proximity layer \textbf{b.} thickness and \textbf{c.} magnetization. In the left panels of \textbf{b} and \textbf{c}, black and red solid lines denote optimal thresholds on the true and predicted values, respectively, which subdivide the histogram into four rectangular quadrants delimiting the false positive (upper left), true positive (upper right), false negative (lower right), and true negative (lower left) examples, where the positive class exhibits proximity magnetism. The right panels of \textbf{b} and \textbf{c} show the true and false positive rates corresponding to different choices of threshold for both the true and predicted parameter values, with each point colored by the predicted threshold value (always the larger of the two). The optimal thresholds are selected to balance the trade-off between true and false positive rates, corresponding to the outlined point in the right panels.}}
    \label{fig:FigS14}
\end{figure*}

\begin{figure*}[t]
    \centering
    \includegraphics[width=0.9\linewidth]{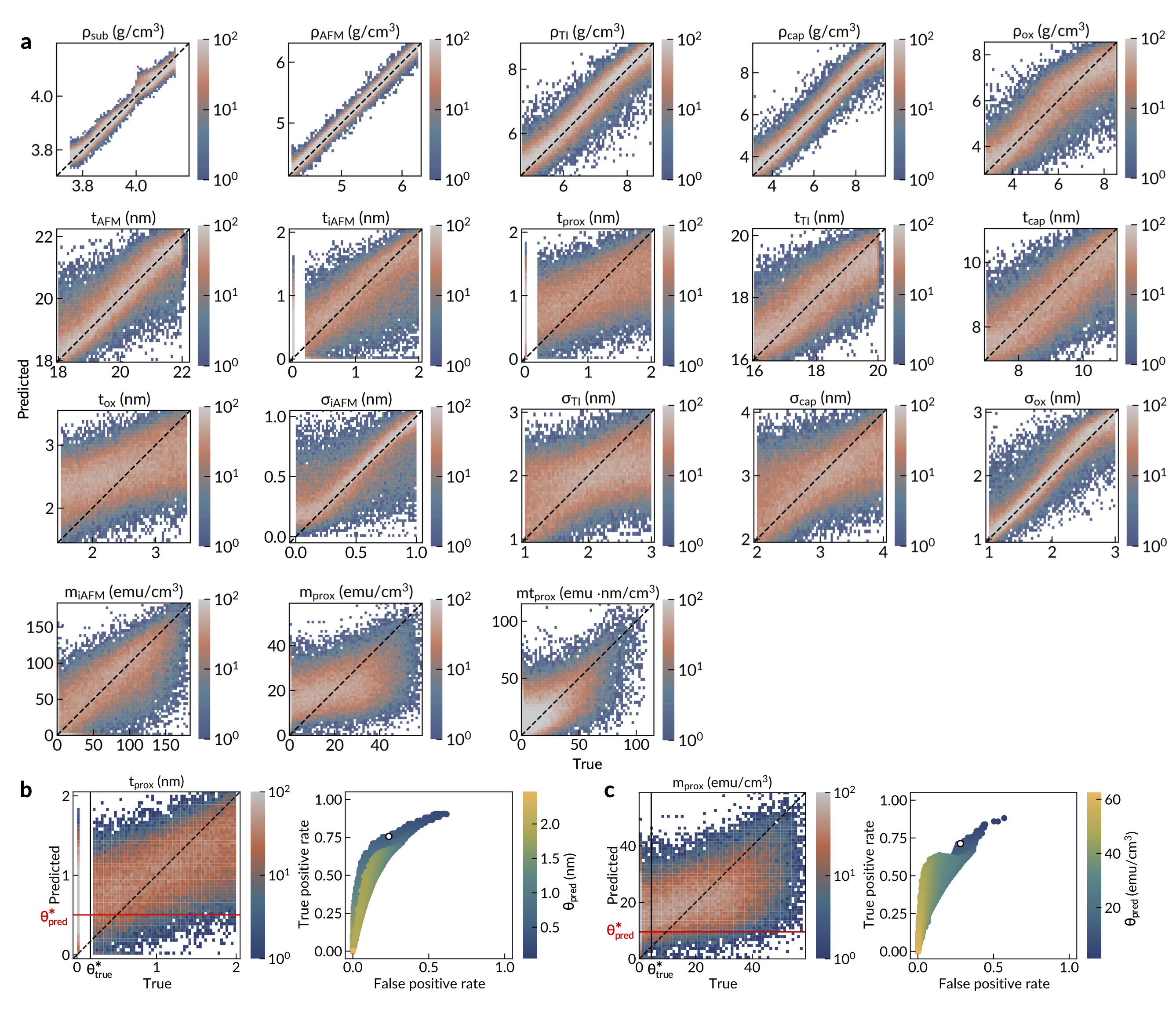}
    \caption{\textbf{Regressor performance for the (Bi,Sb)$_2$Te$_3$/Cr$_2$O$_3$ system.} \textbf{a.} Histograms of predicted versus true values for different sample parameters of the test dataset. {\color{revision}\textbf{b-c.} Threshold determination for proximity layer \textbf{b.} thickness and \textbf{c.} magnetization. In the left panels of \textbf{b} and \textbf{c}, black and red solid lines denote optimal thresholds on the true and predicted values, respectively, which subdivide the histogram into four rectangular quadrants delimiting the false positive (upper left), true positive (upper right), false negative (lower right), and true negative (lower left) examples, where the positive class exhibits proximity magnetism. The right panels of \textbf{b} and \textbf{c} show the true and false positive rates corresponding to different choices of threshold for both the true and predicted parameter values, with each point colored by the predicted threshold value (always the larger of the two). The optimal thresholds are selected to balance the trade-off between true and false positive rates, corresponding to the outlined point in the right panels.}}
    \label{fig:FigS15}
\end{figure*}

\begin{figure*}[h]
    \centering
    \includegraphics[width=0.9\linewidth]{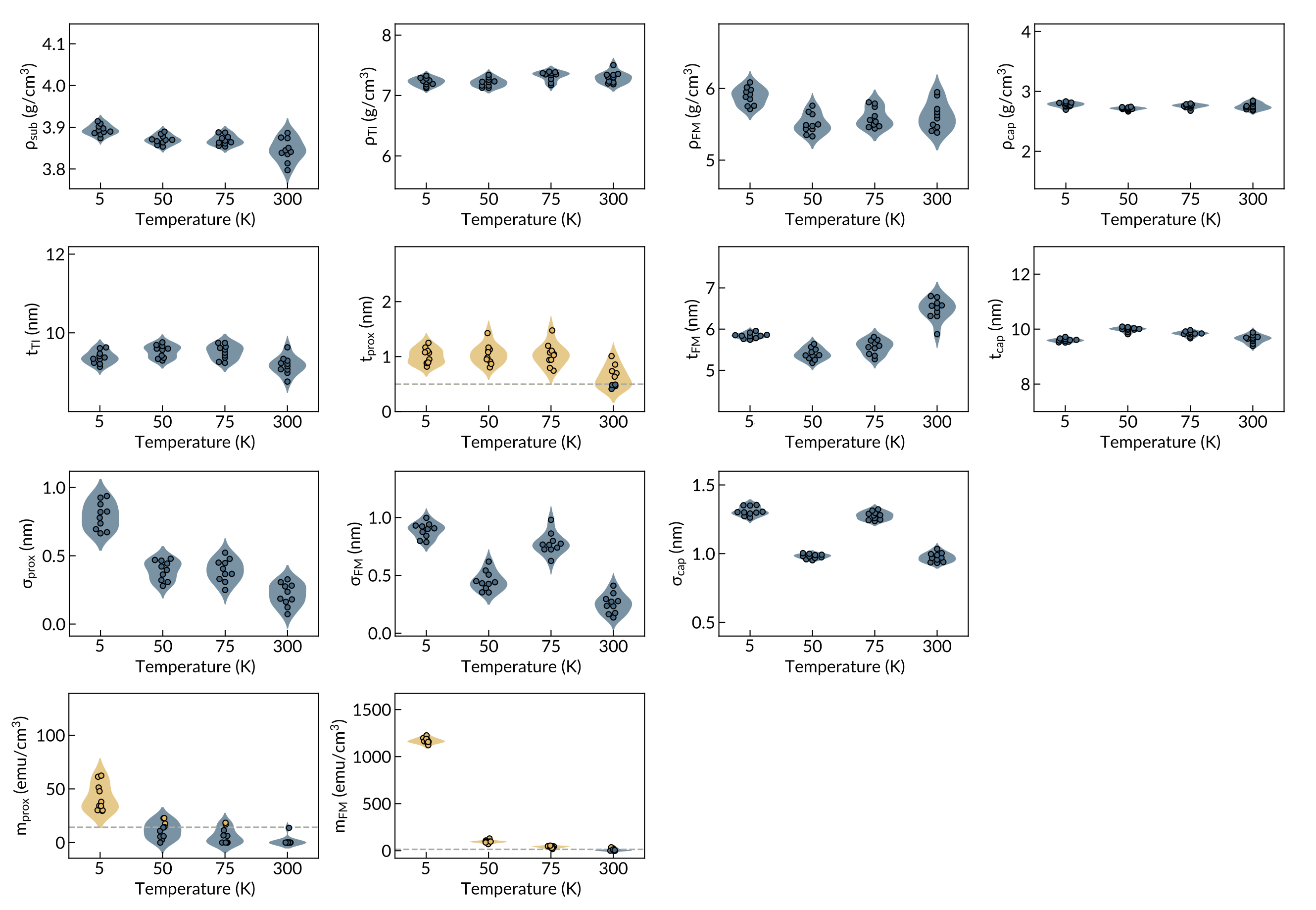}
    \caption{\textbf{Statistics of predicted parameter values for the Bi$_2$Se$_3$/EuS system.} The predictions of each sample parameter obtained from 10 instances of the VAE trained with different initial weights, shown as a function of the measurement temperature of the corresponding experiment. Gray dashed lines indicate the optimal threshold obtained for proximity classification. Scattered points above (below) the determined threshold are colored yellow (blue). Violin plots of the predicted values for experiments with majority predictions above (below) the threshold are shaded yellow (blue).}
    \label{fig:FigS16}
\end{figure*}

\begin{figure*}[t]
    \centering
    \includegraphics[width=0.9\linewidth]{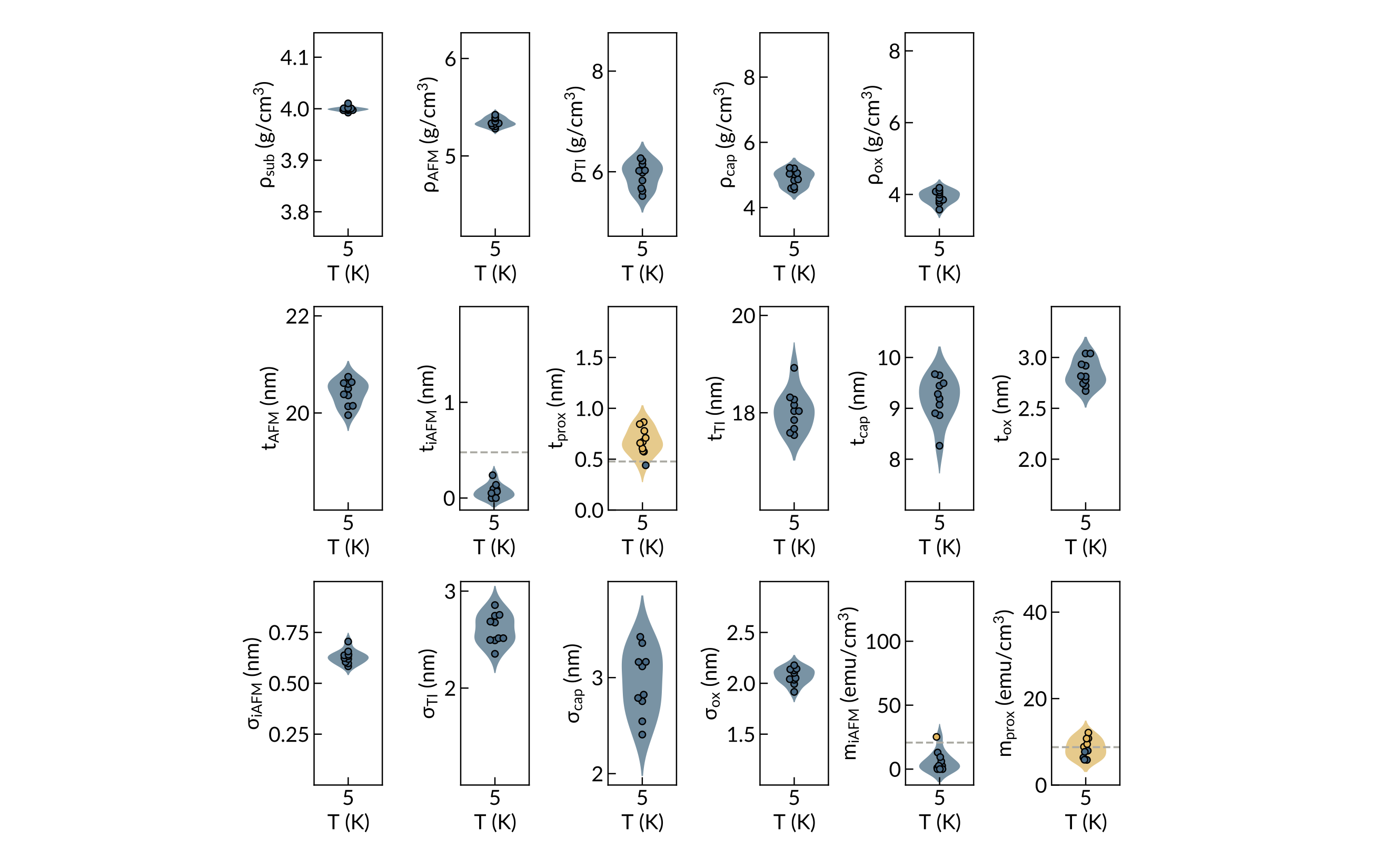}
    \caption{\textbf{Statistics of predicted parameter values for the (Bi,Sb)$_2$Te$_3$/Cr$_2$O$_3$ system.} The predictions of each sample parameter obtained from 10 instances of the VAE trained with different initial weights. Gray dashed lines indicate the optimal threshold obtained for proximity classification. Scattered points above (below) the determined threshold are colored yellow (blue).}
    \label{fig:FigS17}
\end{figure*}

{\color{revision}The complete sets of histograms visualizing the VAE regressor performance for the Bi$_2$Se$_3$/EuS and (Bi,Sb)$_2$Te$_3$/Cr$_2$O$_3$ systems are shown in Figs. \ref{fig:FigS14}a and \ref{fig:FigS15}a, respectively. Additionally, Figs. \ref{fig:FigS14}b-c and \ref{fig:FigS15}b-c illustrate the method of threshold determination used to classify the presence or absence of interfacial TI and AFM layers. In particular, the resolution limits of the trained model in recovering the values of $t_\text{prox}$ and $m_\text{prox}$ (and $t_\text{iAFM}$ and $m_\text{iAFM}$ for the (Bi,Sb)$_2$Te$_3$/Cr$_2$O$_3$ system), are obtained by the following procedure. In each case, we consider a pair of thresholds, $\theta_\text{true}$ and $\theta_\text{pred}$, ($\theta_\text{pred} > th_\text{true}$), corresponding to the true and predicted parameter values, respectively. The true threshold is the theoretical resolution limit set by, for instance, the parameter ranges used to generate the data or the experimental resolution, while the predicted threshold is the resolution limit set by the trained model, \textit{i.e.} it incorporates the errors of imperfect prediction by the VAE regressor. When displayed on the two-dimensional histograms shown in the left panels of Figs. \ref{fig:FigS14}b-c and \ref{fig:FigS15}b-c, these thresholds subdivide the plot into four rectangular quadrants delimiting the false positive (upper left), true positive (upper right), false negative (lower right), and true negative (lower left) examples, where the positive class is the one exhibiting proximity magnetism. By varying the choice of both thresholds over the possible range of parameters, we compute the relationship between the true (tpr) and false (fpr) positive rates, plotted in the right panels of Figs. \ref{fig:FigS14}b-c and \ref{fig:FigS15}b-c. The optimal thresholds can then be chosen to balance the trade-off between tpr and fpr by minimizing the following objective function using the true and predicted parameter values of the validation set:}
\begin{equation*}
    (\theta_\mathrm{true}^*, \theta_\mathrm{pred}^*) = \argmin_{(\theta_\mathrm{true}, \theta_\mathrm{pred})}\Big(\text{fpr}(\theta_\mathrm{true}, \theta_\mathrm{pred}) - \text{tpr}(\theta_\mathrm{true}, \theta_\mathrm{pred}) + \Big| \text{tpr}(\theta_\mathrm{true}, \theta_\mathrm{pred}) - (1 - \text{fpr}(\theta_\mathrm{true}, \theta_\mathrm{pred})) \Big| \Big).
\end{equation*}
{\color{revision}The objective of the above is to obtain the lowest possible fpr with the highest possible tpr while minimizing the discrepancy between tpr and $(1-\text{fpr})$, also known as the true negative rate, ensuring that recovery of both positive and negative classes is comparably accurate. We define the average $\theta_\mathrm{pred}^*$ obtained over 10 instances of the trained model with different initial weights as the resolution limit of the VAE approach for each of $t_\text{prox}$, $m_\text{prox}$, $t_\text{iAFM}$, and $m_\text{iAFM}$, corresponding to the gray dashed lines plotted in Figs. \ref{fig:FigS16} and \ref{fig:FigS17} as described in the main text.}

\clearpage
{\color{revision}\section*{Reproducibility of VAE predictions}}
{\color{revision}We show the predictions for the full set of sample parameters produced by ten trained models with different initial weights for the Bi$_2$Se$_3$/EuS and (Bi,Sb)$_2$Te$_3$/Cr$_2$O$_3$ systems in Figs. \ref{fig:FigS16} and \ref{fig:FigS17}, respectively. In both cases, the predicted values for most parameters are similar across different models, particularly for bulk properties, but a few parameters exhibit greater uncertainty (\textit{e.g.} $\sigma_\text{cap}$ in Fig. \ref{fig:FigS17}. Additionally, while the temperature dependence of the quantities in Fig. \ref{fig:FigS16} is mostly as expected, a few exceptions are worth mentioning. First, the values of most bulk temperature-independent quantities, such as materials' densities and thicknesses, are predicted consistently across different temperatures. However, there is a slight increase in the thickness of the FM layer at 300 K which appears to coincide with slight reductions in $t_\text{TI}$ and $t_\text{prox}$, and a more significant reduction in $\sigma_\text{prox}$. In fact, the predicted roughnesses for all interfaces and surfaces is lowest at 300 K, but reasons for this cannot be fully understood from the present model. Additionally, while $\sigma_\text{prox}$ generally decreases with increased temperature, the roughness values for the FM and capping layers do not appear to follow a clear temperature-dependent trend.}

\end{document}